\documentclass[
reprint,
superscriptaddress,
amsmath,
amssymb,
aps,
prx,
floatfix,
]{revtex4-2}
\usepackage{ragged2e}
\usepackage{caption}
\DeclareCaptionJustification{justified}{\justifying}
\usepackage[colorlinks,allcolors=cyan]{hyperref}
 
 \newcommand{\ptmi}{P\qty[\mathcal{I}_3=0]}
\usepackage{graphicx}
\usepackage{dcolumn}
\usepackage{bm}
\usepackage[capitalise]{cleveref}
\usepackage{physics}
\usepackage{enumitem}
\usepackage{xcolor}
\usepackage{comment}
\usepackage{bbold}
\usepackage{wrapfig}
\usepackage{mathtools}
\usepackage{amsmath}
\usepackage{lipsum}  
\usepackage[caption=false]{subfig}
\DeclareCaptionOption{justified}[]{\caption@setjustification{justified}}

\newif\ifSM
\SMtrue
\newif\ifmainText
\mainTexttrue 

\begin{document}
\def\lc{\left\lfloor}   
\def\rc{\right\rfloor}
\setlength{\intextsep}{10pt plus 2pt minus 2pt}
\setlength{\abovedisplayskip}{4pt}
\setlength{\belowdisplayskip}{4pt}

\ifmainText
\title{Infinitely fast critical dynamics: Teleportation through temporal rare regions in monitored quantum circuits}
 
\author{Gal Shkolnik}
 \affiliation{Racah Institute of Physics, The Hebrew University of Jerusalem, Jerusalem 91904, Israel}
 
 \author{Sarang Gopalakrishnan}
\affiliation{Department of Electrical and Computer Engineering, Princeton University, Princeton, NJ 08544, USA}

 \author{David A. Huse}
 \affiliation{Department of Physics, Princeton University, Princeton, New Jersey 08544, USA}
 
 \author{Snir Gazit}
 \affiliation{Racah Institute of Physics, The Hebrew University of Jerusalem, Jerusalem 91904, Israel}
 \affiliation{The Fritz Haber Research Center for Molecular Dynamics, The Hebrew University of Jerusalem, Jerusalem 91904, Israel}

 \author{J. H. Pixley}
  \affiliation{Department of Physics and Astronomy, Center for Materials Theory,
Rutgers University, Piscataway, New Jersey 08854, USA}
\affiliation{Center for Computational Quantum Physics, Flatiron Institute, New York, New York 10010, USA}

\date{\today}

\begin{abstract}

We consider measurement-induced phase transitions in monitored quantum circuits with a measurement rate that fluctuates in time, remaining spatially uniform at each time. The spatially correlated fluctuations in the measurement rate disrupt the volume-law phase for low measurement rates; at a critical measurement rate, they give rise to an entanglement phase transition with ``ultrafast'' dynamics, i.e., spacetime ($x,t$) scaling $\log x \sim t^{\psi_\tau}$.  The ultrafast dynamics at the critical point can be viewed as a spacetime-rotated version of an infinite-randomness critical point; despite the spatial locality of the dynamics, ultrafast information propagation is possible because of measurement-induced quantum teleportation. We identify temporal Griffiths phases on either side of this critical point. We provide a physical interpretation of these phases, and support it with extensive numerical simulations of information propagation and entanglement dynamics in stabilizer circuits. The implications of our results on the general stability of phase transitions and ordered phases to such temporal randomness
are discussed.

\end{abstract}

\maketitle

\section{Introduction} 
Enriching quantum dynamics beyond the unitary limit through the use of measurements has demonstrated significant potential for quantum state preparation and quantum error correction \cite{gottesman1999, PhysRevLett.86.5188,PhysRevLett.86.910, PhysRevX.14.021040,PhysRevLett.131.200201,lee2022decoding,hong2024quantum}. 
Measurements have two qualitatively different effects on the dynamics of entanglement in quantum many-body systems: measuring a subsystem to be in a unique pure state disentangles it; 
however, such measurements can also \emph{teleport} information in the rest of the system, transmuting short-range entanglement into long-range entanglement \cite{PhysRevLett.70.1895,PhysRevLett.81.5932,PhysRevB.104.104305}. These effects underlie two remarkable information-theoretic phase transitions that were discovered in the past few years. The first is the measurement-induced phase transition (MIPT) in the entanglement structure of monitored quantum circuits, driven by the competition between the entangling power of unitary dynamics and the disentangling effects of measurements \cite{Fisher2018Zeno,Fisher2019MIPT,Nahum2019MIPT,Fisher_2023}; the second is the teleportation transition, at which measuring all but two well-separated spins in a many-body system creates entanglement between the remaining spins \cite{Altman2024finitetime}. Because measurements can teleport information, monitored circuits do not have strict causal light-cones. This apparent lack of causality is consistent with the laws of physics because to send signals through teleportation, one needs to perform classical communication, which cannot be superluminal. It was recently shown \cite{Ultrafast2023} that certain measures of quantum information propagate superballistically (with the dynamical scaling $x \sim t^{3/2}$) in monitored circuits in the low-measurement phase. Nevertheless, the critical point at the ``standard'' MIPT is governed by a conformal field theory, with \emph{emergent} isotropic space-time scaling (i.e. it has a dynamical exponent $z$ that relates space and time scaling as $t\sim x^z$ with $z=1$) \cite{Gullans2020Dynamical,PhysRevB.104.104305,Zabalo-2022}.

\begin{figure}[b!]
\vspace{-2.0mm}
\begin{center}
    \includegraphics[width=0.48\textwidth]{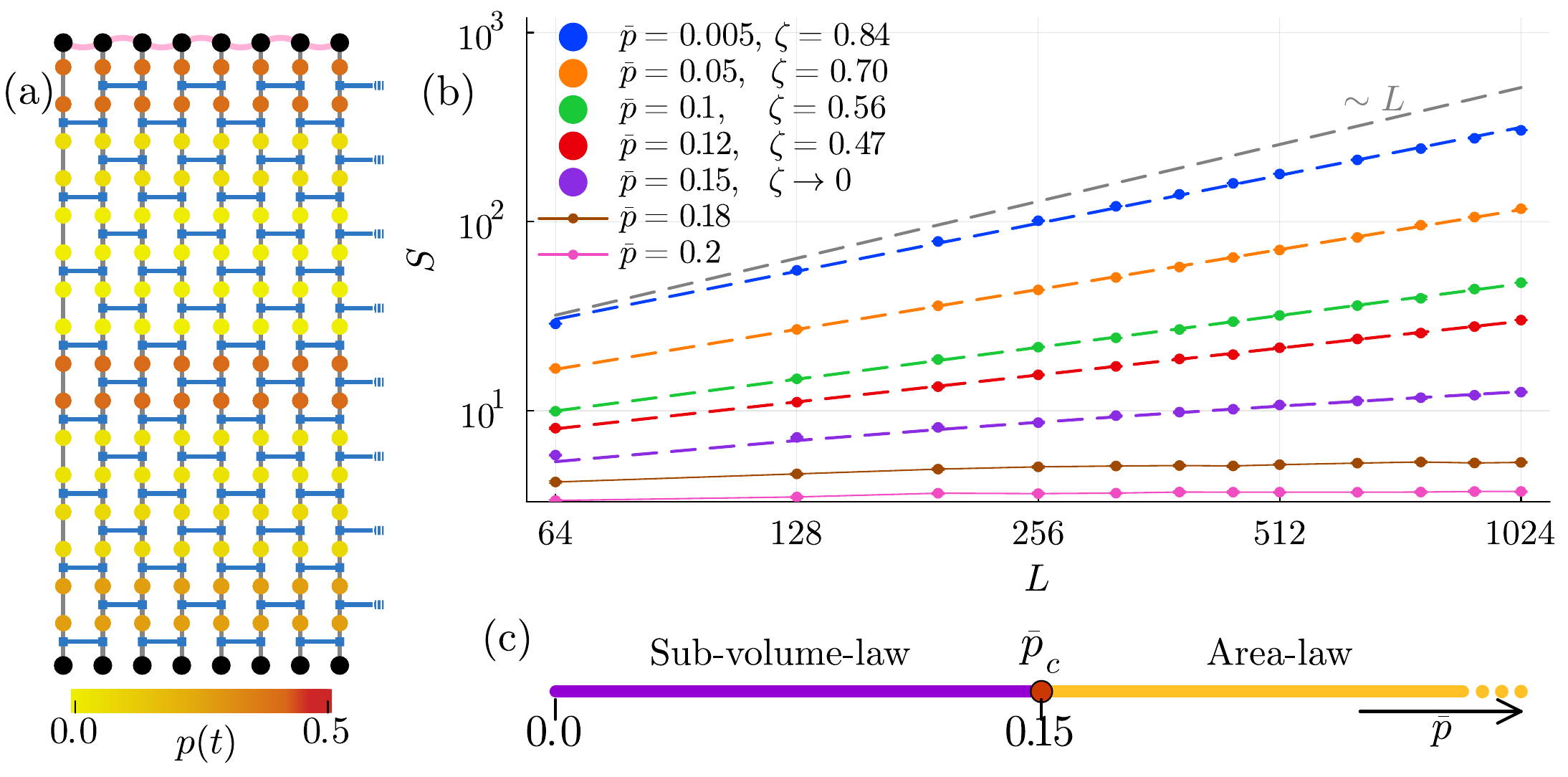}
  \end{center}
  \vspace{0.0mm}
  \captionsetup[subfigure]{labelformat=empty}
    \subfloat[\label{subfig:sys_sketch}]{}
    \subfloat[\label{subfig:S_vs_L}]{}
    \subfloat[\label{subfig:phase_diag}]{}
  \caption{(a) Graphical depiction of a brickwork quantum circuit with temporal randomness in the measurement rate. The blue lines mark the two-qubit Clifford gates, while the projective measurements operating between them follow a time-dependent measurement rate $p(t)$ denoted by the different colors. (b) The steady-state half-cut bipartite entanglement entropy for chains of $L$ qubits with periodic boundary conditions at different values of $\bar{p}$ grows as $S\sim L^{\zeta}$ with $\zeta(\bar{p})<1$ for $0<\bar{p}<\bar{p}_c$, obeying a sub-volume-law behavior. (c) The phase diagram of the temporally random quantum circuit.} 
  \label{fig:fig1}
\end{figure}

Similar but different from MIPTs, the stability of quantum codes to noise gives rise to noise threshold transitions (NTTs) that lie within universality classes distinct from the MIPT \cite{Preskill2002Topo,PhysRevB.109.054432,PhysRevB.109.125148,putz2025flow}. Understanding the stability of quantum codes, and the resulting threshold transitions to different types of disorder and noise that arise physically across the many available noisy intermediate-scale quantum (NISQ) devices \cite{Preskill2018quantumcomputingin,10.5555/2011665.2011666,Preskill2002Topo,PhysRevA.89.022321, PhysRevA.62.062311,hoke2023quantum} remains a central question of both theoretical and practical importance. In solid-state devices (e.g. superconducting qubits), imperfections intrinsic to the materials used represent a source of static disorder in these devices. Work on the MIPT and threshold transitions \cite{qhz8-tvwb} has demonstrated that the power of scaling theory, namely the Harris criteria, can be directly applied to these dynamical transitions in quantum circuits to ascertain their stability, which requires $\nu \ge 2/d$ where $\nu$ is the correlation length exponent (in the absence of static disorder) in $d$ dimensions. In one dimension, this usually leads to an instability driving the critical point to an infinite-randomness fixed point. NISQ devices also have several other forms of disorder that are less commonplace than in conventional condensed matter systems, such as burst errors that produce correlated noise affecting 
a large number of qubits at one time, sometimes across the whole system. This can be driven naturally by ionizing radiation (e.g. cosmic rays such as muons) in superconducting circuits that can activate a large block of contiguous qubits for a short period of time or due to fluctuations in global control protocols \cite{mcewen2022resolving,tan2024resilience,li2025cosmic}. It is therefore imperative to determine the stability criterion for such correlated forms of temporal randomness in non-equilibrium MIPTs and threshold transitions, as well as in phase transitions and ordered phases much more generally.  The effect of such temporal randomness can be considered for any model with temporally sustained dynamics.

In this work, we leverage the teleporting power of measurements to create a critical point that strongly deviates from the relativistic scaling, with instead a vanishing dynamical exponent $z \to 0$, which we will term ultrafast dynamical scaling. The model we consider consists of a random quantum circuit subject to measurements occurring at a rate that fluctuates in time, but in a globally spatially correlated way. Our scheme bears resemblance to the problem of burst errors: there, errors (instead of projective measurements) are fully correlated in space.
In particular, therefore, the actual dynamics is \emph{local} in spacetime (being distinct from non-local gates \cite{PhysRevLett.128.010604,PhysRevLett.128.010603,PhysRevLett.128.010605,PhysRevA.104.062405,10.21468/SciPostPhysCore.5.2.023} and collective measurements~\cite{iadecola2024concomitant} producing $0<z<1$): the only nonlocality comes from global correlations of the measurement rate and from the inherent nonlocality of quantum teleportation. 
This type of criticality is qualitatively different from standard quantum criticality: temporal randomness lacks a natural physical interpretation in ground-state quantum criticality; 
considering real-time dynamics opens the door to new types of critical phenomena. 

The setup we are considering can be regarded as a spacetime-rotated \cite{PRXQuantum.2.040319,PhysRevX.12.011045} version of a more standard infinite-randomness MIPT that was studied in \cite{Zabalo2022InfiniteRand}. In that work, measurement rates varied in space but not in time, giving rise to activated $z \to \infty$ critical dynamical scaling. 
Naively, interchanging the roles of space and time maps the $z \to \infty$ critical point to a $z \to 0$ critical point.
However, as we will discuss, quantities like entanglement and mutual information do not map to natural observables under spacetime rotation; therefore, these quantities exhibit unusual features at the ultrafast critical point that have no direct analog in the spacetime-rotated version. This model therefore allows a direct study of temporal randomness and its effect on the stability of the MIPT.

For critical points with spatial quenched randomness, the Harris and the Chayes-Chayes-Fisher-Spencer (CCFS) bounds on the correlation length exponent are $\nu \ge 2/d$ 
\cite{Harris_1974,luck1993classification,PhysRevLett.57.2999,PhysRevB.109.214209}. Our work extends this notion to entanglement transitions with spatially uniform temporal randomness, with the observation that our estimates for the critical exponent for the correlation time $\tau \sim (\bar{p}-\bar{p}_c)^{-\nu_{\tau}}$ saturate a putative temporal Harris/CCFS bound in $d$ dimensions~\cite{Vojta_Spatiotemporal} 
\begin{equation}
    \nu_{\tau}\equiv \nu z \ge 2~.
    \label{eq:harris_tau}
\end{equation}
We note that in the above setting, we consider randomness only along the temporal direction, and hence the inequality is $d$-independent.

Given the numerical estimates of the standard MIPT critical exponents, $\nu=1.28(2)$ and $z=1$ in one dimension \cite{Fisher2019MIPT,Nahum2019MIPT,Fisher_2023,Gullans_2020,Gullans2020Dynamical}, we expect and find that this critical point is unstable to temporal randomness.

Thinking about this bound much more generally: as with the usual (spatial) Harris/CCFS bound, the known critical points populate both sides of this bound with many examples.  For critical points with $\nu z<2$, added spatially uniform temporal randomness is relevant.  In cases where the distinction between the two phases is not stable to such randomness, this could remove the critical point altogether.  But if the distinction between the phases does survive, then the critical point should become some different universality class, and this may be an infinite-randomness class in some cases, as in the model we study here.

In addition to unusual critical behavior, this correlated-measurement model has unusual entanglement properties at low measurement rate.
Chaotic systems are generally expected to exhibit steady-state entanglement that follows a volume-law scaling in the low-measurement phase \cite{PhysRevE.99.032111,PhysRevLett.125.180604,PhysRevB.103.104206}. Nevertheless, recent work has demonstrated that introducing measurements with rates that vary randomly in time gives rise to a plethora of nontrivial scaling laws governing the entanglement entropy, ranging from the standard volume law to sublinear power law, all the way to logarithmic \cite{PhysRevX.12.011045}.

Consistent with these findings, we show that the spatially correlated fluctuations in the rate of single-qubit projective measurements disrupt the volume-law entanglement entropy at low measurement rates. This is due to temporal strips of high measurement rate that interrupt the growth of entanglement, so the stationary entanglement scales as a continuously varying power of subsystem size; this was termed 
the fractal entangled phase by \cite{PhysRevX.12.011045}. This ``interrupted volume law'' phase presents other surprising features: for example, the entanglement across a given cut exhibits temporal fluctuations that are \emph{of the same magnitude} as its average value. These are Griffiths effects, but here due to rare times instead of rare spatial regions.  The area-law phase also has temporal Griffiths effects producing rare times at which some long-range entanglement appears.

This paper is organized as follows: In \cref{sec:model}, we present the model under investigation:  hybrid quantum circuits with temporally random measurement rates. We also detail physical probes employed to identify and characterize the distinct phases and associated phase transition. Moreover, we propose an ultrafast space-time scaling ansatz characterized by activated scaling and a vanishing dynamical exponent $z \rightarrow 0$ that governs the critical dynamics close to the MIPT. In \cref{sec:results}, we detail our numerical results: i) We report unconventional scaling properties of the entangling and area-law phases, in the respective limits of weak and strong average measurement rates.  ii) We conduct a careful analysis of the emergent universal critical properties of the MIPT, demonstrating an ultrafast dynamical scaling and numerical estimation of the associated critical exponents. iii) We examine the signatures of our findings in the context of information propagation and compare our results with the conventional relativistic MIPT and the ultraslow infinite-randomness fixed point in the presence of quenched randomness. \cref{sec:discussion} summarizes our results, discusses their broader implications, and highlights future research directions.

\section{Monitored quantum circuit with temporal randomness}
\label{sec:model}
\subsection{Model}

\begin{figure}[htb!]
\vspace{-8.0mm}
\begin{center}
    \includegraphics[width=0.32\textwidth,angle=270]{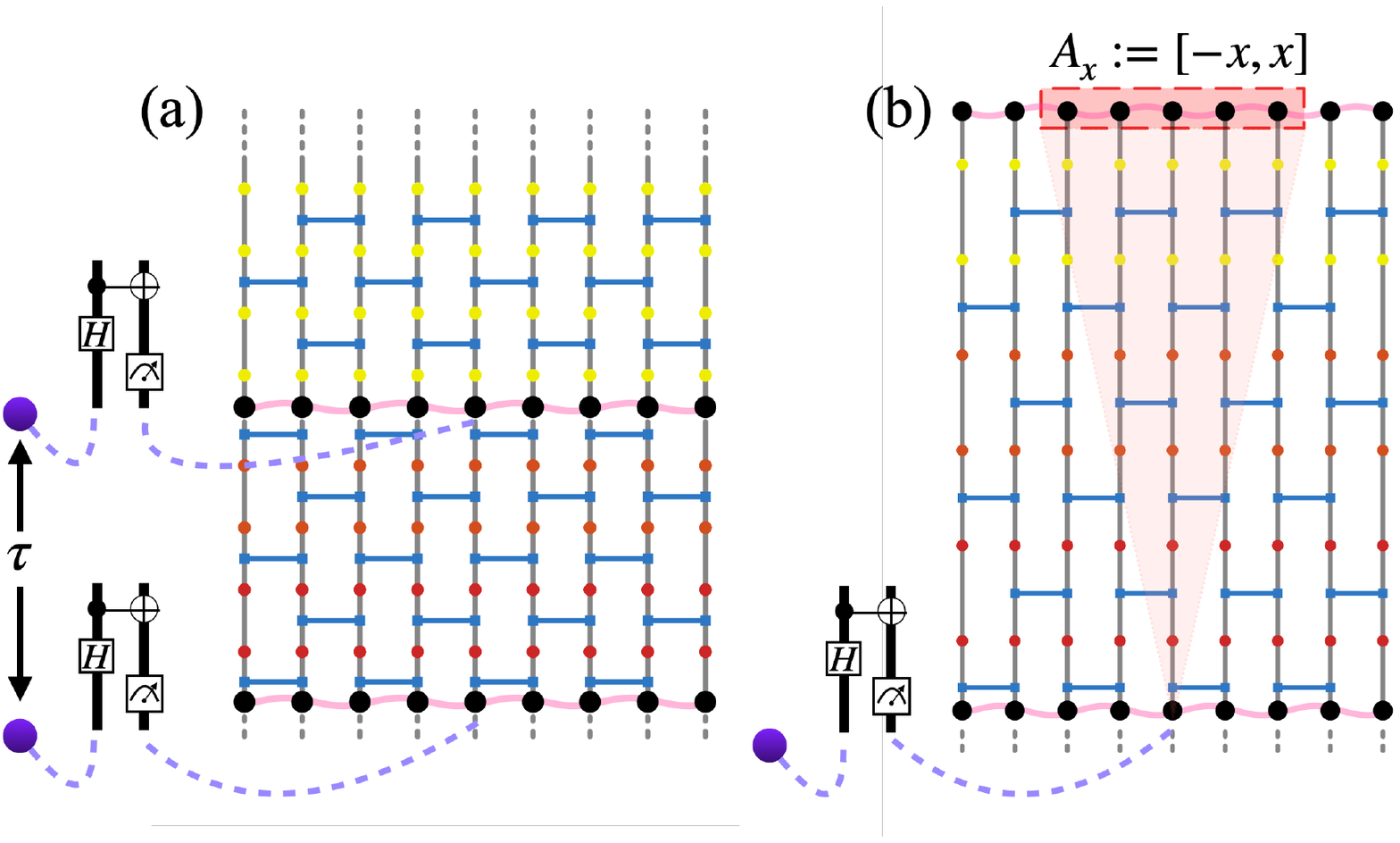}
  \end{center}
  \vspace{-4.0mm}
  \captionsetup[subfigure]{labelformat=empty}
    \subfloat[\label{subfig:two_anc}]{}
    \subfloat[\label{subfig:single_anc}]{}
  \caption{Schematic description of the ancilla probes that we use. In both cases, we first let the circuit evolve until reaching a stationary state. In (a), we measure a qubit in the system and 
  then arrange it in a Bell state with an ancilla qubit. We then let the system evolve for a time $\tau$, and repeat this process with the same site and a second ancilla. We follow by calculating the mutual information between the two ancillas $\mathcal{I}_2(Q_1,Q_2)$ over time. In (b), we use a system with open boundaries.  At the initial time we entangle the ancilla $Q$ with the qubit at $x=0$.
  At later times, we calculate the ancilla's mutual information with a symmetric segment $A_x=[-x,x]$. The information propagation is evaluated using the smallest $x$ for which $\mathcal{I}_2(Q,A_x)>0$.}
  \vspace{-2.0mm}
  \label{fig:fig_ancillas_method}
\end{figure}

To construct the model, we start from the previously-studied model with quenched randomness in the measurement rates
\cite{Zabalo2022InfiniteRand}.  That model had random two-qubit stabilizer gates with local random projective measurements in the computational basis with local measurement rates $p(x)$ that are static in time and only random in space.  We then apply a spacetime rotation to the pattern of measurement rates ($x\rightarrow t$ and $t\rightarrow -x$), which yields measurement rates that are now uniform in space but random in time, $p(x)\rightarrow p(t)$, as depicted in \cref{subfig:sys_sketch}.
This is not a full rotation of the previous model, because we are not rotating the gates.
We note that we could, in principle, take dual-unitary Clifford gates \cite{PhysRevLett.123.210601,10.21468/SciPostPhys.8.4.067} that are in the same universality class as random Cliffords \cite{Zabalo-2022,Aziz-2024,Kumar-2024}, and they would be invariant under this rotation.  So we expect that in their bulks these two models related by this spacetime rotation are in the same universality class governed by an infinite-randomness fixed point. 
The rare regions in space that dominate the dynamics at the infinite-randomness fixed point now become rare regions in time, and below we unveil unique aspects of such temporal Griffith effects. Relatedly, temporal Griffiths phases were recently reported in the context of decodable phases of error correction schemes in the presence of non-uniform noise rates \cite{qhz8-tvwb}, and in classical dynamical models \cite{PhysRevE.85.051125,JJAlonso_2001}.

Specifically, at each time $t$, we draw the measurement rate across the entire lattice according to the random variable $p(t)=\frac{1}{2}r_t^n$, where $r_t$ are independent random numbers uniformly distributed on $[0,1]$. With this choice, we control the time-averaged measurement rate $\bar{p}$ by varying $n$, with $\bar{p}=1/\qty(2n+2)$. 
We note that the limit $n\to0$ corresponds to a deterministic (time-independent) measurement rate $p(t)=\bar{p}|_{n\to0}=1/2>p_c^{\text{\tiny MIPT}}\approx0.16$ that places the model in the area-law phase \cite{Fisher2019MIPT,Nahum2019MIPT,Gullans2020Dynamical}. Whereas, for $n\to\infty$, the measurement rate drops to zero, resulting in unitary dynamics and a volume-law phase. Dynamic rare time intervals appear from the measurement rate associated with a \emph{specific} time being greater or lesser than $p_c^{\text{\tiny MIPT}}$. We explore the MIPT in this spatially uniform model by varying $n$ and thus $\bar{p}$. Unless stated otherwise, we take periodic boundary conditions. 

\subsection{Entanglement probes}
 
The main quantities we study are obtained from the bipartite entanglement entropy defined as 
\begin{equation}
    S(A)=-\text{Tr}_A\qty[\rho_A\log_2\qty(\rho_A)]~,
\end{equation} 
with the reduced density matrix $\rho_A=\text{Tr}_{A^C}\ket{\psi}\bra{\psi}$, for a spatial partition of a segment $A$ and its complement $A^C$, and the full system in pure state $|\psi\rangle$. With the above definition, we can construct the bipartite and tripartite mutual information defined as 
\begin{equation}
    \mathcal{I}_2(A,B)\coloneqq S(A)+S(B)-S(A\cup B)
\end{equation} 
and
\begin{equation}
\mathcal{I}_3(A,B,C)\coloneqq\mathcal{I}_2(A,B)+\mathcal{I}_2(A,C)-\mathcal{I}_2(A,B\cup C)~,
\end{equation} respectively. These latter two quantities can be made particularly useful by choosing appropriate geometries and state preparations with ancillas.

To quantify the scaling of space-time dynamics, we use an ancilla qubit that we entangle with the monitored quantum circuit in two different ways. First, following the protocol of \cite{Gullans_2020,Zabalo2022InfiniteRand} we track the purification time of the ancilla. To do so, we start by unitarily evolving the system for a duration of $T=2L$, and then we generate a Bell pair between a system qubit and an ancilla qubit. We then let the circuit evolve according to the model described above and calculate the bipartite entanglement entropy $S_Q(t)$ of the ancilla qubit with the rest of the system after time $t$. Second, we use the mutual information between two different ancillas. To that end, we first let the system reach a stationary state, then we measure a qubit in the system and put it in a Bell state with 
an ancilla $Q_1$; after $\tau$ timesteps, we repeat this process with a second ancilla $Q_2$ and the same qubit.  Then we continue running the circuit while calculating the mutual information between the two ancillas at each timestep, 
$\mathcal{I}_2(Q_1,Q_2;t)$, see \cref{subfig:two_anc}.

\begin{figure*}[th!]
\begin{center}
    \includegraphics[width=\textwidth]{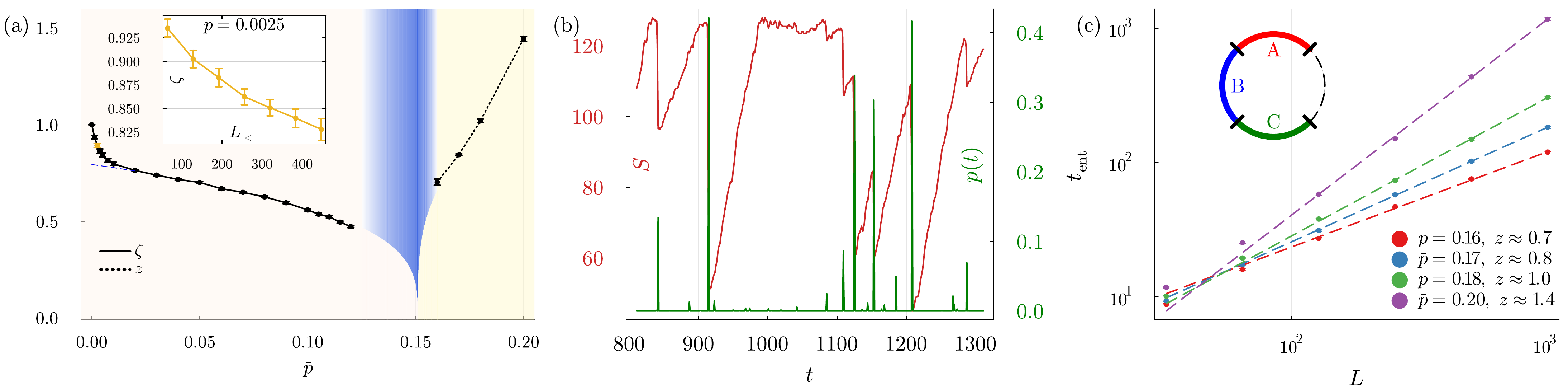}
  \end{center}
  \vspace{0.0mm}
  \captionsetup[subfigure]{labelformat=empty}
    \subfloat[\label{subfig:zeta_grif}]{}
    \subfloat[\label{subfig:sawtooth}]{}
    \subfloat[\label{subfig:z_area}]{}
  \caption{ (a) The exponent $\zeta(\bar{p})$ (solid black line), controlling the saturated entanglement entropy scaling $S\sim L^\zeta$, continuously decreases with an increase in $\bar{p}$ in the entangling phase. Our finite-size numerical estimates suggest $\zeta<1$ even for the smallest positive $\bar{p}$, see the decreasing exponent shown in the inset. The dashed blue line serves as an estimate of thermodynamic values in the low $\bar{p}$ regime, calculated by a power-law fit to the data points at sufficiently large $\bar{p}$, which converged to the thermodynamics limit within our finite-size data. For $\bar{p}>\bar{p}_c$, we depict the dynamical exponent $z$ (dashed black line), evaluated via the entanglement time, $t_{\text{ent}}$, which decreases upon approach to the critical point. 
  (b) Temporal sawtooth behavior of the half-cut bipartite entanglement entropy $S$ for a specific circuit realization. Here $L=256$ and $\bar{p}=0.005$. The green line marks the measurement rate at time $t$, $p(t)$. Each sharp drop in $S$ coincides with a significant peak of $p(t)$ corresponding to a high measurement rate. The sawtooth shape gradually disappears upon approaching criticality. 
  (c) Numerical estimates of the dynamical exponent $z$ in the area-law phase, calculated using finite-size scaling of the entanglement time of an initial product state, defined as the first time for which $\mathcal{I}_3(A,B,C)\neq 0$ for any possible partition of the system in to adjacent segments $\{A,B,C\}$ each of length $L/4$; an example for such a partition is depicted in the inset.
  }
  \vspace{2.0mm}
  \label{fig:fig3}
\end{figure*}

Last, to understand the consequences of this ultrafast dynamics, we monitor the speed at which quantum information spreads, similarly to the scheme presented in \cite{Ultrafast2023}. Explicitly, we consider open boundary conditions and let the system evolve for a duration $\tau_0$ beyond the saturation time of the average entanglement entropy. Next, we carry out a projective measure on the central qubit and entangle it with a single ancilla qubit to form a Bell pair. We reset our clock to $t=0$ and proceed with the circuit dynamics. At each time instance, as long as the ancilla is still entangled with the system, we calculate the evolution of the bipartite mutual information $\mathcal{I}_2(Q,A_x)$ between the ancilla $Q$ and a symmetric segment of the system around the central qubit $A_x\coloneqq[-x,x]$, see \cref{subfig:single_anc}. We denote by $x_I(t)$ the smallest $x$ for which $\mathcal{I}_2(Q,A_x)$ is nonvanishing. This quantity serves as a measure of the minimal segment size that allows us to infer information encoded in the ancilla qubit. In the presence of entangling dynamics, for a given segment size $x$ smaller than the full system, information on the ancilla qubit will (provided entanglement with the system remains) typically leak outside the segment, so the mean $x_I(t)$ will grow with time and thus give an estimate of the dynamics of information spreading.

Unless specified otherwise, all quantities considered are averaged.  Each quantity is averaged over different realizations of the quantum circuit, the initial product state, and the random structure of the measurement rate.  Typically, we use at least 3000 different realizations to ensure convergence.

\subsection{Infinitely Fast Scaling Ansatz}
To guide our exploration, we postulate the anticipated critical space-time scaling of our dynamics by drawing inspiration from the dual problem of quenched, i.e., time-invariant spatial modulations, either random or quasiperiodic \cite{Zabalo2022InfiniteRand,Shkolnik2023QP}. There, the critical dynamics present ultraslow activated scaling of time scales versus the system size $\log t \sim L^\psi$, with an activation exponent $\psi$, similarly to the infinite-randomness criticality of random one-dimensional spin chains \cite{Fisher1992TFIM,Young_1996,Senthil1996randomPotts,PhysRevB.104.214208}. This motivates exchanging the roles of space and time, which suggests the following \emph{ultrafast} critical scaling hypothesis
\begin{equation}
    t^{\psi_{\tau}} \sim \log L
    \label{eq:t_L_scaling}
\end{equation}
for a temporal critical activation exponent $\psi_{\tau}$. In the following, we will test the above scaling ansatz in our model by examining the scaling properties of various quantities.

To facilitate comparison between the spatially random and temporally random models, Table~\ref{tab:spacetime_rotation} summarizes the key observables and their corresponding quantities under the spacetime rotation transformation.

\begingroup
\squeezetable
\begin{table}[h]
\begin{ruledtabular}
\begin{tabular}{ccc}
Spatial randomness & Temporal randomness & Physical meaning \\ \hline
$\xi$ & $\tau$ & \parbox[t]{0.28\columnwidth}{Correlation length/time} \\ \hline
$\nu$ & $\nu_\tau=\nu z$ & \parbox[t]{0.28\columnwidth}{Correlation length/time exponent} \\ \hline
$\psi$\quad $\log t \sim L^{\psi}$ & $\psi_\tau$\quad $\log L \sim t^{\psi_{\tau}}$ & \parbox[t]{0.28\columnwidth}{Activation exponent of stretched exponential / logarithmic space-time scaling} \\ \hline
Diverging $z$\quad $z=\frac{L^{\psi}}{\log L}$ & Vanishing $z$\quad $z=\frac{\log \log L}{\psi_\tau \log L}$ & \parbox[t]{0.28\columnwidth}{Dynamical exponent limit behavior} \\
\end{tabular}
\end{ruledtabular}
\vspace{10.0mm}
\caption{Spacetime rotation correspondence between spatially random and temporally random models.}
\label{tab:spacetime_rotation}
\end{table}
\endgroup

\section{Results}
\label{sec:results}
\subsection{Sub-volume law entangling phase}

We initiate our analysis by studying the stability of the volume-law phase to the introduction of a time-varying random measurement rate corresponding to a small nonzero $\bar{p}$. To begin,
we examine the deviations from the volume-law scaling of the long-time entanglement entropy, $S\sim L$, that is present at $\bar{p}=0$.
To that end, we compute the growth exponent associated with the power law ansatz, compatible with the fractal entanglement phase in Ref.~\cite{PhysRevX.12.011045},
\begin{equation}
    S(L,t\to \infty) \sim L^{\zeta(\bar{p})}~;
\label{eqn:subvolumelaw}
\end{equation} 
see \cref{subfig:S_vs_L}. Here, $S(L,t)$ denotes the half-cut bipartite entanglement entropy taken $t$ time-steps after the initial state, and the limit $t\to \infty$ corresponds to times greater than the saturation time of the averaged entanglement entropy. 
Within the resolution limit of our numerics, we find that the volume-law phase appears to be unstable, with a sudden drop in $\zeta(\bar{p})$ below unity, even for a small 
$\bar{p}>0$.  Further increasing $\bar{p}$ continuously decreases $\zeta$ upon approaching the MIPT.
We interpret, and provide direct evidence below, that these results can be understood through a Griffiths effect in time, which leads to a sublinear scaling of the entanglement entropy with system size in the entangling phase, see \cref{subfig:zeta_grif}. Further increasing $\bar{p}$ beyond the critical value $\bar{p}_c$ completely suppresses the growth of the entanglement entropy, leading to an area-law phase.

As additional support, we also examine the 
bipartite entanglement entropy at late times as a function of the subsystem length $x$ and find 
a sublinear growth with an exponent $\zeta$, $S(x;L,t\rightarrow\infty)\sim x^{\zeta(\bar{p})}$ (for $x\ll L$) that agrees well with the previous analysis. As before, ${\zeta(\bar{p})}$ continuously decreases with increasing $\bar{p}$, see \cref{apdx:Sx}. This sublinear growth of the late-time entanglement with $x$ is the spacetime rotation of the sublinear-in-time growth of entanglement in the model with quenched randomness \cite{Zabalo2022InfiniteRand}, so is thus as expected. 

The sub-linear growth of the long-time entanglement with $L$ is a natural consequence of temporal rare region effects. To illustrate the origin of this phenomenon, we present in \cref{subfig:sawtooth} a typical time trace of the entanglement entropy $S(t)$ for a specific random circuit realization, characterized by a relatively small average measurement rate ($\bar{p}=0.005$).  With the above choice, for the most part $p(t)$ is extremely small, interleaved by rare time instances that have a high measurement rate. The resulting temporal evolution follows a ``sawtooth'' structure. 
Specifically, $p(t)$ times with high measurement rate give rise to an abrupt drop in the entanglement entropy, followed by a recovery period, during which $S$ grows linearly with time, attempting to exhaust its maximum value $L/2$. We emphasize that the maximal value of $S$ is $L$-dependent, whereas the rate of occurrence of large $p(t)$ events is $L$-independent. The competition between the above two mechanisms is at the core of the nontrivial $L$-dependence of the average long-time entanglement entropy. For small $\bar{p}$ values, this behavior and the sublinear growth are captured by an effective heuristic model, presented in \cref{apdx:sawtooth_model}. 

It is also interesting to understand the temporal growth of the entanglement entropy in the sub-volume-law phase. For the spacetime dual model with quenched randomness, the long-time saturated entanglement entropy was assumed to grow linearly (``volume law") with the system size in Ref.~\cite{Zabalo2020Critical}. This scaling translates under spacetime rotations to linear-in-time growth of the entanglement entropy in the thermodynamic limit. In \cref{apdx:dynamical_exps}, we test the above prediction. Curiously, our finite-size simulations are consistent with a sub-linear temporal growth. Assessing the stability of these results in the thermodynamic limit requires considering simulations of system sizes beyond our numerical capabilities.

\subsection{Area law phase}

Generally, a high measurement rate can contribute to teleportation and allow the information to spread super-ballistically; the presence of the measurements means that Lieb-Robinson bounds \cite{lieb1972finite,Ultrafast2023} do not apply. However, too many measurements also suppress the entanglement and may restrict its spreading. We hypothesize that the critical point is exactly where the interplay between teleportation and entanglement suppression leads to the fastest growth of the entanglement range. Therefore, we expect the dynamical exponent $z$, relating the scales of space and time, to be smallest at $\bar{p}_c$ and grow again upon increasing the measurement rate. 

To study the behavior of the dynamical exponent in the area-law phase, we look for the first time in which long-range entanglement emerges. Concretely, we initialize the system in a random product state and monitor the growth of entanglement by tracking the tripartite mutual information, $\mathcal{I}_3(A,B,C)$, across all possible partitions $\{A,B,C\}$ of the system, where $A,B,C$ are adjacent quarters. We identify the first time when $\mathcal{I}_3(A,B,C)\neq 0$, serving as a typical time for the emergence of long-range entanglement, and examine how this time scales with the system size $L$, positing an $t_{\text{ent}}\sim L^z$ scaling relation. We note that this time is associated with rare events that become increasingly rare with the increase in $\bar{p}$. This long-range entanglement does not persist in the area-law phase; it is produced at time $t_{\text{ent}}$ by a low-measurement temporal rare event but is then quickly removed by subsequent measurements.

Indeed, as we show in \cref{subfig:z_area}, the exponent $z$ grows with increasing $\bar{p}$ in the area-law phase. Overall, completing the picture of minimal, vanishing $z$ at $\bar{p}_c$ and a temporal Griffiths fan opening from both sides, as presented by a blue shade in \cref{subfig:zeta_grif}.

\begin{figure*}[ht!]
\begin{center}
    \includegraphics[width=1.0\textwidth]{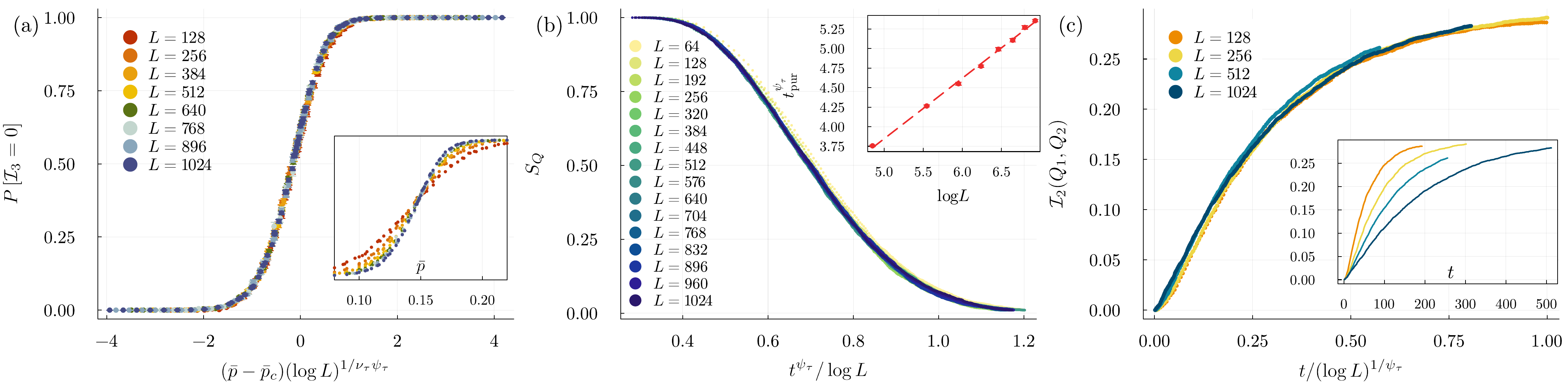}
  \end{center}
  \vspace{-0.0mm}
  \captionsetup[subfigure]{labelformat=empty}
    \subfloat[\label{subfig:ptmi_collapse}]{}
    \subfloat[\label{subfig:anc_collapse}]{}
    \subfloat[\label{subfig:corrT_collapse}]{}
  \caption{(a) Curve collapse analysis of the universal amplitude $\ptmi$, according to the ansatz in \cref{eqn:ptmi_ansatz}, yields a horizontal axis scaling with the values $\bar{p}_c=0.151$ and $\nu_{\tau}\psi_{\tau}=0.6$. The unscaled data is depicted in the inset. (b) The single ancilla entanglement $S_Q(t)$ and (c) the mutual information between two ancillas separated by $\tau=16$ scale as functions of $t^{\psi_{\tau}}/\log L$ with $\psi_{\tau}=0.3$ at $\bar{p}=0.151$. The inset in (b) shows that the same scaling applies to the average ancilla purification time, namely the average time it takes an ancilla to disentangle from the system completely.}
  \label{fig:fig4}
\end{figure*}

\subsection{Infinitely fast fixed point}

Motivated by the above findings, we now turn to study the phase transition separating the sub-volume- and area-law phases. To determine the critical properties, we postulate the emergence of a diverging time scale $\tau$ (i.e. a correlation time) close to criticality, with a power-law scaling 
\begin{equation}
    \tau\sim |\bar{p}-\bar{p}_c|^{-\nu_\tau},
    \label{eqn:correlation-time}
\end{equation} 
for a correlation time exponent $\nu_\tau$. We choose to work with this definition as it is the spacetime rotation of the scaling of the correlation length for the system with quenched disorder.  If the scaling for our temporally-random model were all power laws it would be
$\xi\sim |\bar p-\bar p_c|^{-\nu}$ for the correlation \emph{length} and $\tau \sim \xi ^z  \sim |\bar p-\bar p_c|^{-\nu z}$. The implied relation based on the above is $\nu_\tau=\nu z$. However, we expect ultrafast critical dynamics, i.e., $z\rightarrow 0$ and hence $\nu \rightarrow \infty$ such that $\nu_\tau$ is a constant (as we show below). 

Due to temporal rare regions, the distribution of $P[\mathcal{I}_3]$ is broad, see \cref{apdx:I3_dist}, and a similar approach to that used for the infinite-randomness transition can be employed based on $\ptmi$ \cite{Zabalo2022InfiniteRand}: In the sub-volume law entangling phase we will almost always find $\mathcal{I}_3\neq 0$, while in the area-law phase the probability that $\mathcal{I}_3=0$ becomes unity.
Using the scaling from \cref{eq:t_L_scaling}, we obtain a generalized finite-size scaling ansatz for the 
probability $\ptmi$ \cite{Gullans2020Dynamical,Zabalo2020Critical,Zabalo2022InfiniteRand}
\begin{equation}
    \ptmi(\bar{p},L)\sim f\qty[\qty(\bar{p}-\bar{p}_c)\qty(\log L)^{1/\qty(\nu_{\tau}\psi_{\tau})}],
\label{eqn:ptmi_ansatz}
\end{equation}
Here, $f(x)$ is a universal scaling function. 

We employ the standard finite-size scaling curve collapse analysis according to \cref{eqn:ptmi_ansatz} to provide a numerical estimate of the critical average measurement rate $\bar{p}_c=0.150(3)$ and the product of critical exponents $\nu_{\tau}\psi_{\tau}=0.58(2)$, see \cref{subfig:ptmi_collapse} and additional analysis in \cref{apdx:criticality}. Crucially, the critical properties do not follow the scaling relation associated with the spacetime uniform MIPT \cite{Nahum2019MIPT,Fisher2019MIPT,Gullans2020Dynamical}. In particular, attempting a power-law (instead of activated) space-time scaling fails to produce an accurate curve collapse and results in the anticipated running exponents with system size, see \cref{apdx:criticality}. 

\begin{figure}[hb!]
\begin{center}
    \includegraphics[width=0.48\textwidth]{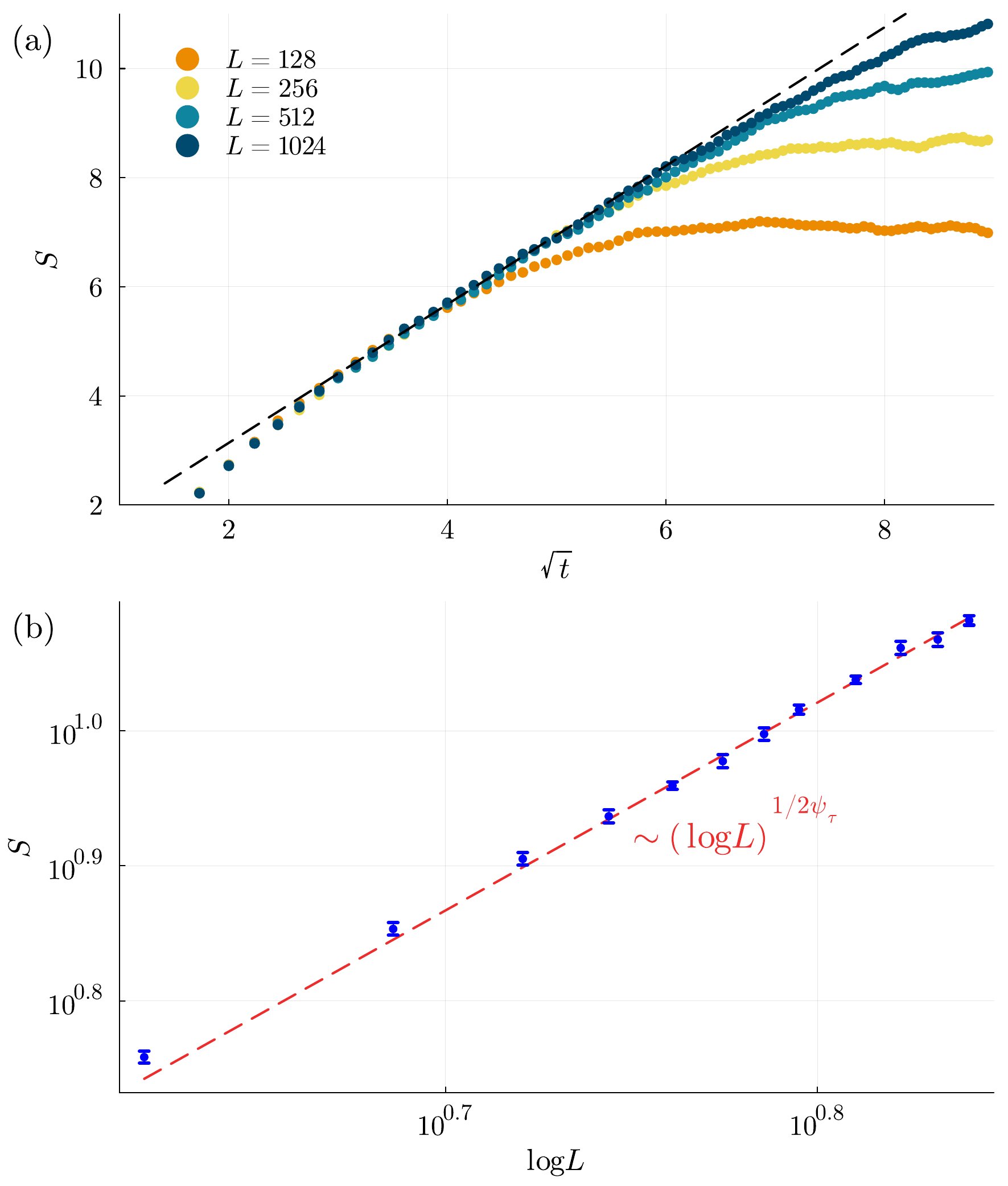}
  \end{center}
  \vspace{2.0mm}
  \captionsetup[subfigure]{labelformat=empty}
    \subfloat[\label{subfig:Spc_vs_t}]{}
    \subfloat[\label{subfig:Spc_vs_logL}]{}
  \caption{(a) The early time entanglement entropy before saturation grows at the critical point as $S\sim t^{1/2}$, a direct rotation of the infinite-randomness model. (b) Combined with the ultrafast scaling, we derive the system size dependence of the late-time saturated entanglement entropy at $\bar{p}_c$: $S\sim \qty(\log L)^{1/2\psi_{\tau}}$, with $\psi_{\tau}=0.32(3)$ over the range studied.}
  \label{fig:fig5}
\end{figure}

As a complementary analysis, allowing us to extract an estimate of $\psi_{\tau}$ directly (without the multiplicative factor $\nu_\tau$), we study the critical dynamics of an ancilla qubit at criticality $\bar{p}=\bar{p}_c$.
We employ the space-time scaling ansatz \cref{eq:t_L_scaling} on the time evolution of $S_Q$ for varying system sizes by presenting the temporal axis in terms of the scaling variable $t^{\psi_{\tau}}/\log L$, see \cref{subfig:anc_collapse}. We obtain a curve collapse for $\psi_{\tau}=0.30(2)$. Our error estimate includes the uncertainty in computing $\bar{p}_c$. Combining the above two results for the critical exponents, we can numerically estimate the correlation time exponent, defined in Eq.~\eqref{eqn:correlation-time}, to be $\nu_{\tau}\approx 1.9(2)$, thus saturating (a spacetime rotated) Harris/CCFS bound within the error bars, as was seen for the rotated model in \cite{Zabalo2022InfiniteRand,Shkolnik2023QP}. A similar estimate for $\psi_{\tau}=0.30(2)$ was obtained using the purification time, defined as the time at which the ancilla qubit disentangles from the system, as shown in the inset of \cref{subfig:anc_collapse}. This key finding provides direct evidence for the ultra-fast dynamical scaling of \cref{eq:t_L_scaling}. We further corroborate this result using the mutual information between two ancillas separated by time $\tau=16$, as described in \cref{subfig:two_anc}, and in \cref{subfig:corrT_collapse} we show that, indeed, the data scales well for different system sizes.

Employed with this definition of the temporal critical exponents, we find our results are consistent with a dynamical Harris criterion. Namely that the infinitely fast fixed point flows to a value that saturates the Harris/CCFS bound for disordered systems, with its suitable spacetime rotation, leads us to the general dynamical hypothesis
$
    \nu_\tau \ge 2
$, \cref{eq:harris_tau}
which our results saturate.

In that regard, Ref.~\cite{Vojta_Spatiotemporal} proposed a complementary bound, $z\nu \ge 2$, relating the dynamical exponent $z$ and the spatial correlation length $\nu$. This entails a divergence of the correlation length exponent, $\nu\to\infty$, upon approaching our critical point with $z\to0$  and $\nu_\tau=\nu z=1.9 (2)$ goes to a finite value saturating the bound.

It is interesting to understand the universal growth of the entanglement entropy at criticality. We begin by drawing inspiration from the spacetime-rotated infinite-randomness problem \cite{Zabalo2022InfiniteRand}, for which $S(L,t\rightarrow \infty)\sim L^{1/2}$ and $S(L\rightarrow\infty,t)\sim \log t$. By combining this perspective with the space-time scaling of \cref{eq:t_L_scaling}, we postulate that critical entanglement evolution should follow an ultrafast scaling form 
\begin{eqnarray}
   &S(L\rightarrow\infty,t)\sim \sqrt{t},
   \\
    &S(L,t\rightarrow \infty)\sim (\log L)^{\frac{1}{2\psi_\tau}}
    \label{eqn:S_scaling}
\end{eqnarray}
We find both of these forms to be consistent with our numerical results, as shown in \cref{fig:fig5}. We caution that precise numerical estimates of these exponents are challenging to extract due to our limited time window before saturation in our finite-size simulations, so our estimates might contain large finite-time and/or finite-size effects.  On the other hand, for the finite-size scaling, we find excellent agreement with the above scaling form.  One should be similarly cautious about the exponents estimated in \cite{Zabalo2022InfiniteRand} for the model with quenched randomness.  Thus the difference between our $\psi_\tau\cong 0.3$ and the $\psi\cong 0.5$ of \cite{Zabalo2022InfiniteRand}, which are expected to be the same by spacetime rotation, may be only due to different finite-size and finite-time effects in the respective data and analyses.

\begin{figure*}[htb!]
\vspace{-0.0mm}
\begin{center}
    \includegraphics[width=1.0\textwidth]{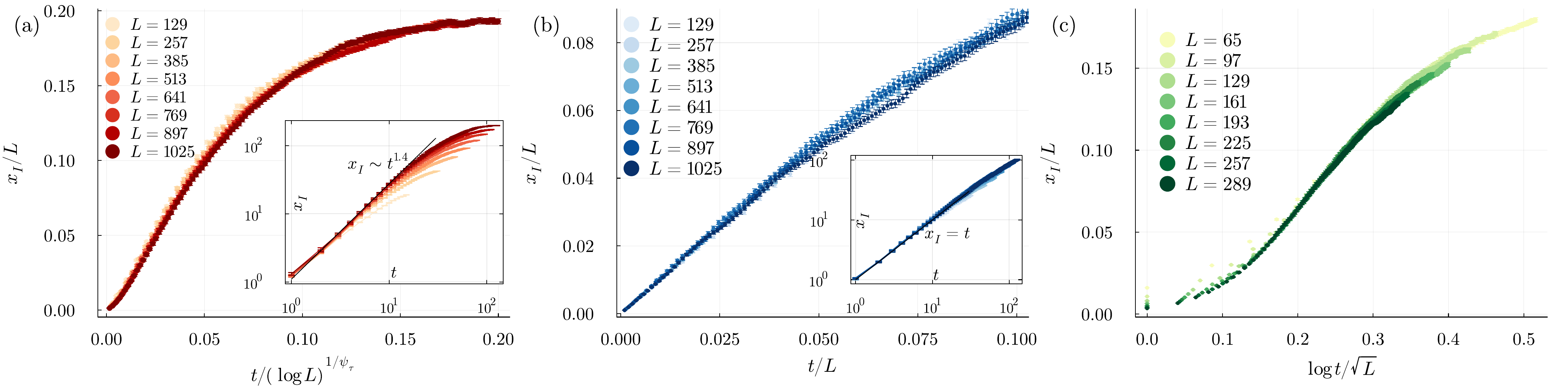}
  \end{center}
  \vspace{-0.0mm}
  \captionsetup[subfigure]{labelformat=empty}
    \subfloat[\label{subfig:xI_dis}]{}
    \subfloat[\label{subfig:xI_uni}]{}
    \subfloat[\label{subfig:xI_infrand}]{}
\caption{(a) At the critical point of the temporally disordered model, the information distance, divided by the linear system size, $x_I/L$, is consistent with the ultrafast scaling ansatz (\cref{eq:t_L_scaling}), with $\psi_{\tau}=0.3$. The inset shows a super-ballistic early-time propagation $x_I\sim t^{\alpha}$ with $\alpha\approx 1.4$. (b) The information distance evaluated for the space-time uniform MIPT displays scaling with dynamical exponent $z=1$. 
The information spreads linearly with time with a velocity of $v=1$, as shown in the inset. (c) At the critical point for the quenched-random model, the information distance follows an ultraslow activated dynamics $x_I/L=f\qty(\log t/\sqrt{L})$.
  }
  \label{fig:fig6}
\end{figure*}

A generalized Widom scaling curve collapse analysis is given in \cref{apdx:criticality}. It is useful to contrast this evolution with the logarithmic growth in space and time at the Lorentz invariant MIPT. 

To further validate our findings and test whether the temporal randomness model truly represents a spacetime rotation of the infinite-randomness fixed point, we studied a ``true'' spacetime-rotation of the quenched disorder model from Ref.~\cite{Zabalo2022InfiniteRand}, see \cref{apdx:direct_rot}. In this complementary approach, we implemented the spacetime rotation at the level of the circuit contraction rather than the measurement pattern. This direct rotation yields critical exponents consistent with our temporal randomness results, with the same ultrafast scaling ansatz and critical behavior. This agreement between two distinct macroscopic implementations, namely, temporal randomness and sideways contraction of spatial quenched randomness, provides strong evidence that we have revealed a robust universality class characterized by ultrafast dynamics with $z\to 0$, rather than merely a phenomenological rotation of the infinite-randomness fixed point.

\subsection{Information Propagation}
Given all of the evidence of the infinitely fast fixed point, we now turn to implications for quantum information propagation.
We study the pace of the information propagation with $x_I(t)$, as defined above, see also \cref{subfig:single_anc}. In \cref{subfig:xI_dis}, we depict $x_I(t)/L$ for a range of systems sizes, at criticality. Plotting the temporal axis with the scaling variable $t/(\log L)^{1/\psi_{\tau}}$ using the previously estimated $\psi_\tau=0.3$, leads to a curve collapse, as before, which nontrivially supports our ultrafast scaling ansatz.

Moreover, we explore the early time dynamics of information spreading at criticality, as shown in the inset of \cref{subfig:xI_dis}. We observe a super-ballistic growth, where $x_I(t)\sim t^\alpha$ with $\alpha\approx1.4$ in this early time range; we expect that this is just an early time effective exponent which would grow if these data could be obtained for much larger system sizes, since the ultrafast scaling has lengths growing faster than any power of time.  This result should be contrasted with the early-time relativistic dynamics of the spacetime uniform MIPT, for which $x_I^{\text{MIPT}}(t) = t$ as directly presented in \cref{subfig:xI_uni}.

To complete the picture and classify the rate of information propagation in the three classes of fixed points we have considered in this work (that have $z\rightarrow 0$, $z=1$, and $z\rightarrow \infty$), we compute $x_I$ also at the infinite-randomness criticality (in the model in Ref.~\cite{Zabalo2022InfiniteRand}). There, the activated scaling takes place with $x_I/L$ being a function of the scaled space-time variable $\log t/\sqrt{L}$, see \cref{subfig:xI_infrand}. This ultraslow dynamics also leaves its mark on our computational efforts and limits us to relatively small system sizes, which makes the determination of the power, or perhaps logarithm, with which $x_I$ grows especially challenging with time. Overall, introducing this correlated randomness to the system allows us to vary the system from ultraslow to ultrafast information propagation.

Lastly, we note that $x_I$ allows us to directly observe teleportation events, through jumps at instants in time that take place in individual samples and are at the core of the superluminal behavior we have uncovered. In \cref{apdx:xI_teleportation}, we demonstrate this effect in a specific circuit realization where teleportation manifests in an abrupt jump in $x_I$.

\section{Discussion}
\label{sec:discussion}

In this work, we have examined hybrid circuit dynamics that follow an inherently non-stationary protocol, realized via a time-dependent and random average measurement rate. The associated circuit dynamics present a competition between time instances that experience few measurements and thus more entanglement production versus other times with many measurements and thus entanglement suppression but more teleportation.

Notably, the resulting emergent phenomena give rise to an unusual entangling phase characterized by a sub-volume growth of the entanglement entropy. In agreement with the fractal entanglement phase \cite{PhysRevX.12.011045}, we understand this behavior as a Griffiths-like effect dominated by strong measurement events that suppress the entanglement entropy even for weak $\bar{p}$. 
This model of temporal noise is relevant to describe burst errors in NISQ devices as arise in ionizing radiation and global control protocols \cite{tan2024resilience, mcewen2022resolving}.
We further show that the transition from this unique phase to the area-law phase displays nontrivial scaling relations that qualitatively match the spacetime-rotated infinite-randomness picture. A striking feature is that this criticality is characterized by ultrafast dynamics with a vanishing dynamical exponent $ z\to 0$. This basic building block allows us to numerically estimate the relevant universal properties, including scaling functions and critical exponents associated with the entanglement entropy growth, temporal fluctuations, and information spreading.

Our results show that the correlation time critical exponent $\nu_{\tau}\cong 2.0$, suggesting a temporal Harris/CCFS bound whose limit is saturated by this model. 
First, we discuss the implications of this result for the MIPT in $d$ dimensions. As demonstrated here the MIPT is unstable in one dimension and flows to an infinitely fast fixed point. In two dimensions $\nu z\approx 0.85$ \cite{PhysRevB.104.155111}, and $\nu$ is expected to keep decreasing as the dimensionality $d$ is increased. Therefore, the MIPT critical point is unstable to temporal randomness also for $d\geq1$.

We can generalize this result beyond MIPTs by applying this stability criterion to NTTs. For example, the threshold transition of two-dimensional surface codes, in the presence of syndrome measurement errors can be mapped to the Nishimori criticality in the 
random-plaquette $\mathbb{Z}_2$ gauge model (RPGM) with $\nu\approx 1.0$. Similarly, the threshold transition in one-dimensional repetition codes is mapped to the phase transition in the random-bond Ising model (RBIM)
where $\nu\approx 3/2$ \cite{WANG200331,Preskill2002Topo}.  
Since in both cases $\nu\leq 2$ and $z=1$, these critical points are unstable to added temporal randomness.
By contrast, the charge-sharpening transition in 1+1-dimensional $U(1)$-conserving random Haar circuits, which has $z=1$ and $\nu\rightarrow\infty$ \cite{PhysRevX.12.041002,PhysRevLett.129.120604,PhysRevB.110.045135}, is stable to added temporal randomness.

Looking to the future, it would be interesting to further engineer the properties of the dynamics by varying the properties of the temporal modulations, e.g. by considering hyperuniform distributions or quasiperiodic modulations. This can potentially allow us to control the activation exponent $\psi_\tau$ and associated critical exponents \cite{Shkolnik2023QP}. Of specific importance would be employing the infinitely fast properties uncovered in this model in the engineering of efficient quantum algorithms and state preparation on quantum computers through the use of feedback based on the measurement outcomes. 

\acknowledgements{
We thank Michael Gullans, Romain Vasseur, Justin Wilson for insightful discussions and collaborations on related work. 
G.S. acknowledges the support of the Council for Higher Education Scholarships Program for Outstanding Doctoral Students in Quantum Science and Technology.
Computational resources were provided by the Intel Labs Academic Compute Environment.
This work is supported in part by the BSF Grant No. 2020264 (G.S., S.Ga., J.H.P.), the Army Research Office Grant No.~W911NF-23-1-0144 (J.H.P.), and U.S. NSF QLCI grant OMA-2120757 (D.A.H.).
}

\appendix

\section{ Entanglement entropy scaling with partition segment size }
\label{apdx:Sx}

As shown in the main text, in the entangling phase, the finite-size scaling of half-cut bipartite entanglement entropy displays a sublinear scaling, deviating from the standard linear growth. Here, we further support this observation via a complementary analysis by tracking the evolution of the entanglement entropy as a function of the partition segment length $x$. In a standard volume-law phase, $S(x)$ is anticipated to grow linearly with $x$ (up to logarithmic corrections) \cite{Fisher2019MIPT}. By contrast, in agreement with our previous results, for temporally random models, we find a sublinear scaling $S(x)\sim x^{\zeta(\bar{p})}$, as shown \cref{fig:supp_Sx}. Crucially, the numerical estimates of $\zeta(\bar{p})$ match the finite-size scaling of the half-cut entropy \cref{subfig:S_vs_L}.

\begin{figure}[h!]
\begin{center}
    \includegraphics[width=0.48\textwidth]{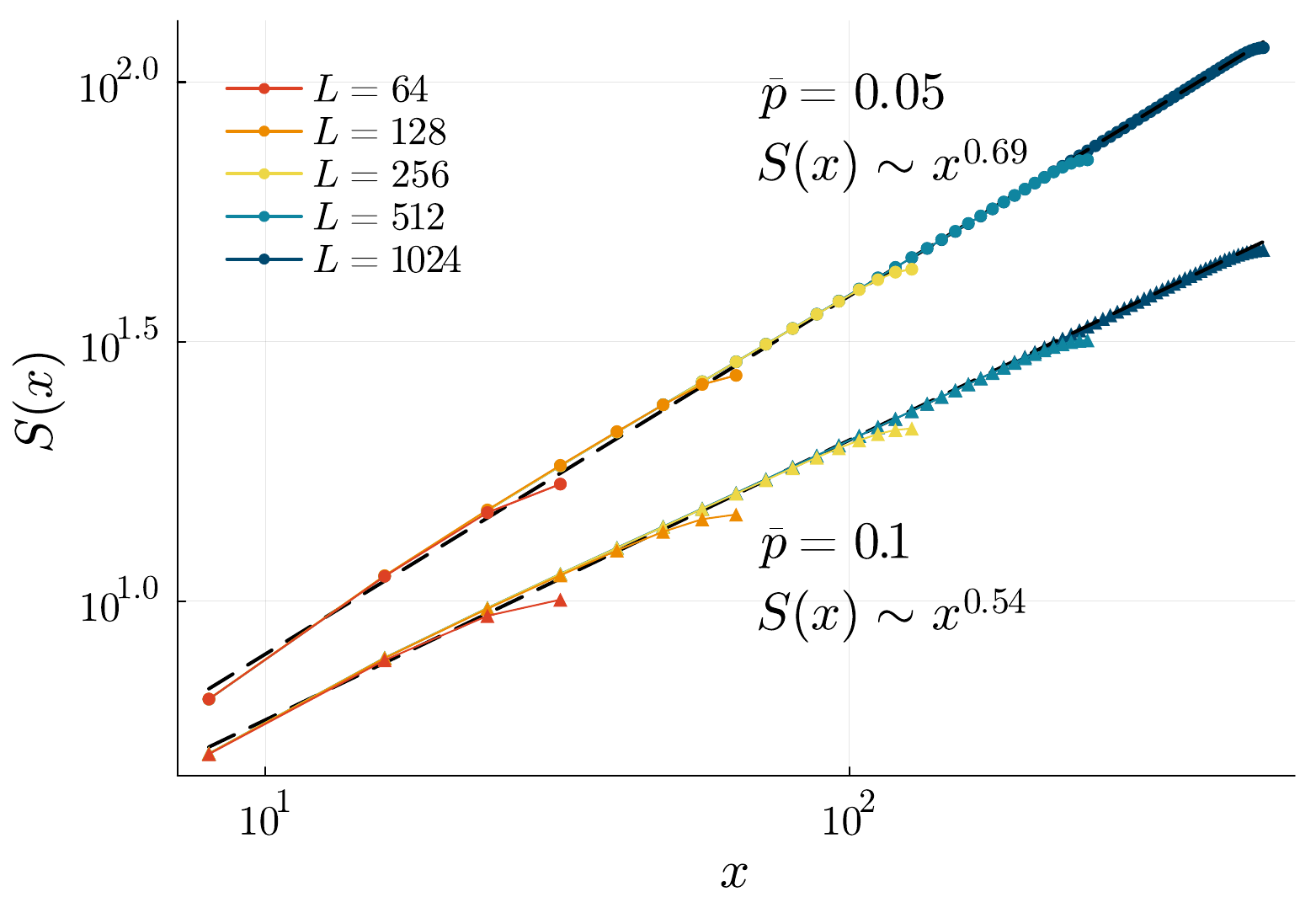}
  \end{center}
  \vspace{4.0mm}
  \caption{The entanglement entropy versus the partition length $x$, in the entangling phase, for a fixed $\bar{p}=0.05,0.1$. Different curves correspond to different system sizes. Dashed lines are a fit to a power law form.}
  \vspace{0.0mm}
  \label{fig:supp_Sx}
\end{figure}

\section{Dynamics of the entanglement entropy in the entangling phase}
\label{apdx:sawtooth_model}

\begin{figure}[bh!]
\begin{center}
    \includegraphics[width=0.48\textwidth]{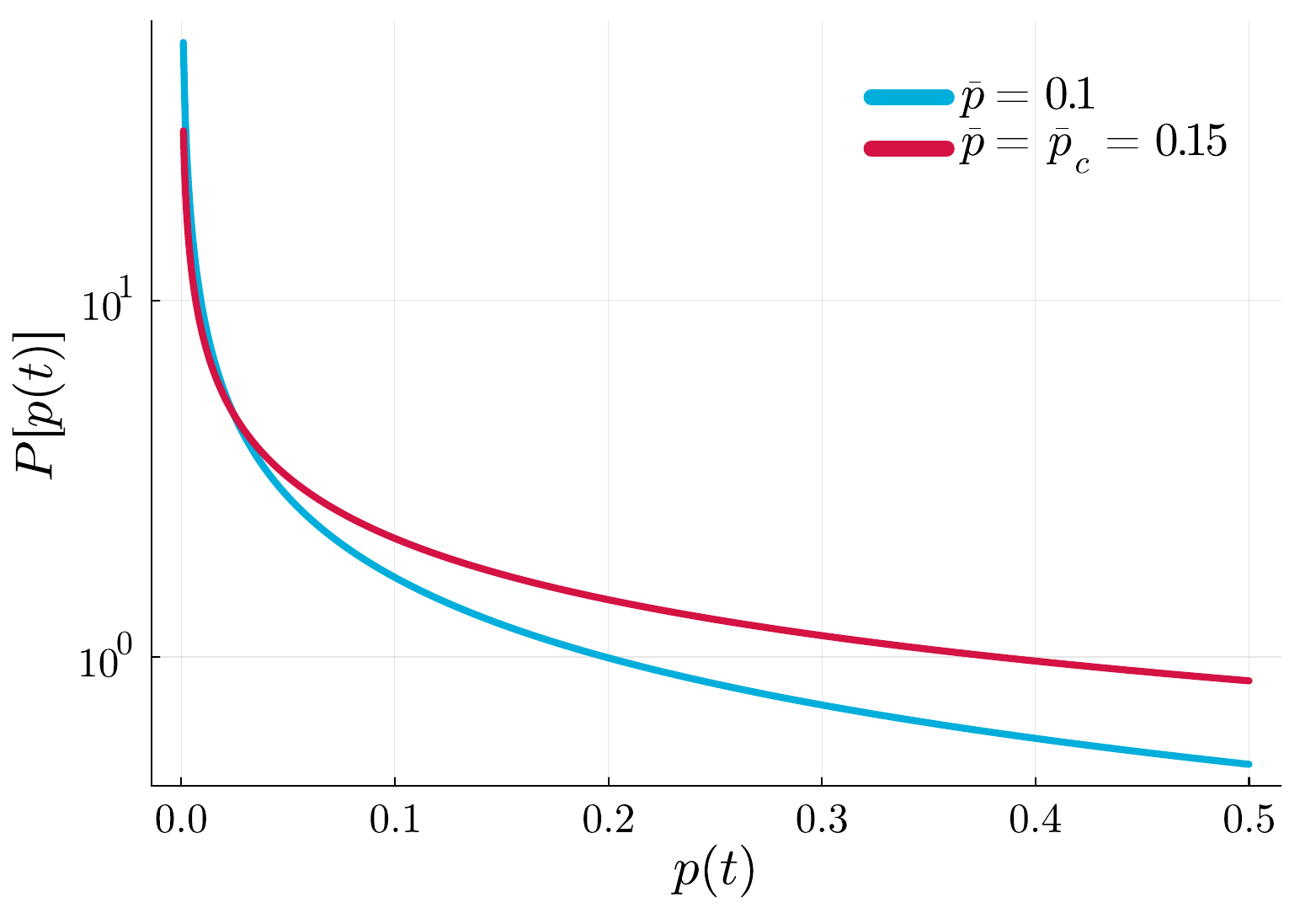}
  \end{center}
  \vspace{3.0mm}
  \caption{The temporal randomness probability distribution function, $p(t)$, with $\bar{p}=0.1$ and $\bar{p}=\bar{p}_c=0.15$. Most events are with a very small $p(t)$, but a long tail of high $p(t)$ values appears.}
  \label{fig:modulation_tail}
\end{figure}

In this section, we analyze the dynamical properties of the entanglement entropy in the entangling phase. 
By design, the temporal distribution is constructed such that for small $\bar{p}$, most outcomes of $p(t)$ are significantly smaller than the critical measurement rate $p_c^{\text{\tiny  MIPT}}$ belonging to the uniform MIPT model. Nevertheless, it still allows rare events with $p(t)$ values corresponding to strong measurement rates. For instance, the probability distribution functions $p(t)$ with $\bar{p}=0.1$ and $\bar{p}=\bar{p}_c=0.15$ are shown \cref{fig:modulation_tail}.

\begin{figure}[h!]
\begin{center}
    \includegraphics[width=0.48\textwidth]{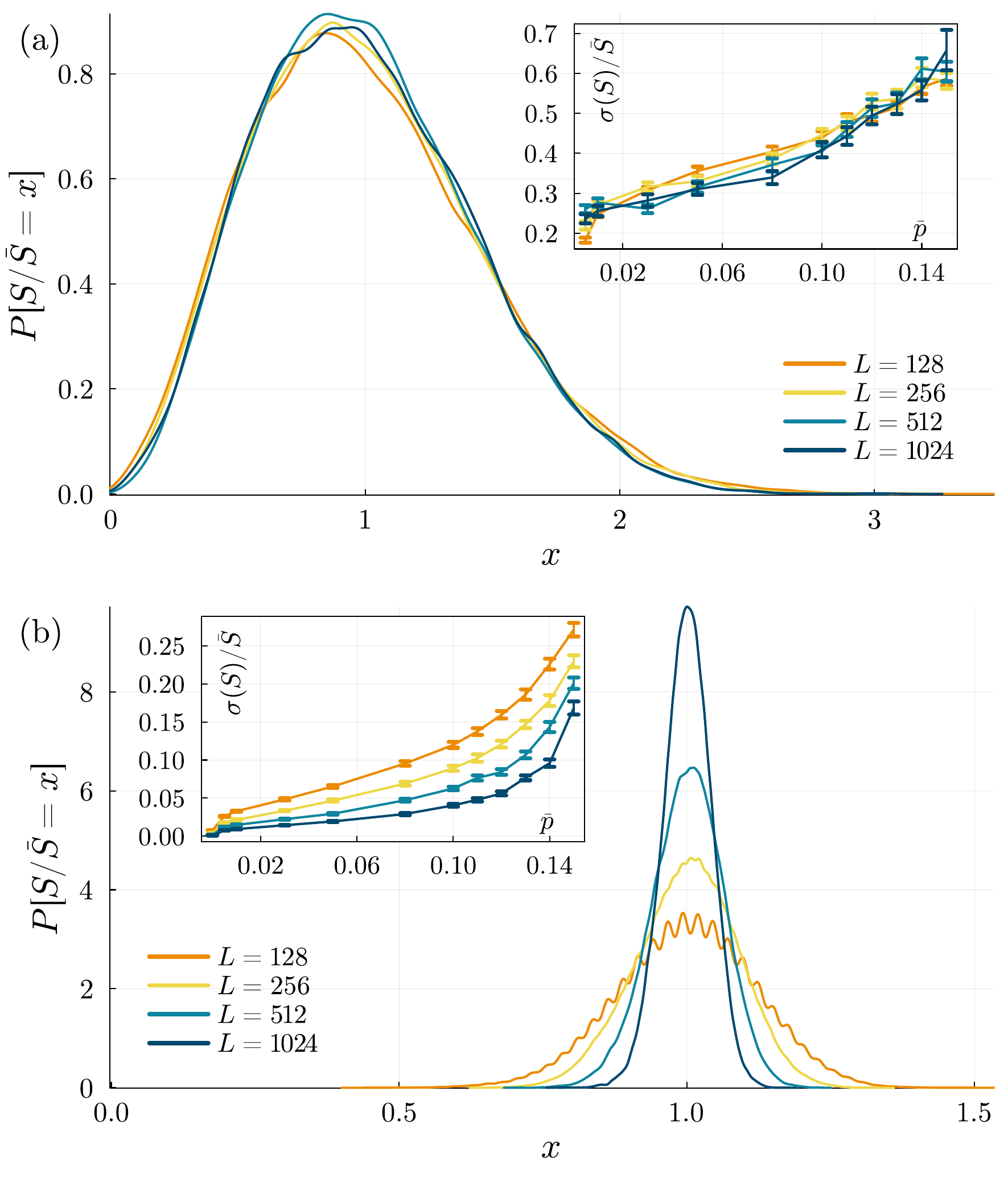}
  \end{center}
  \vspace{5.0mm}
  \caption{The distribution of the entanglement entropy divided by its mean value, $S/\bar{S}$ for $\bar{p}=0.1$ in the (a) temporally random model and (b) the uniform model. In the temporally random model, the distribution is invariant with system size. This implies that the standard deviation of $S$ scales as the average value $\bar{S}$, which is also shown in the inset. In the uniform model, the distribution converges to a deterministic function, meaning that the standard deviation of $S$ scales slower with $L$ compared to the average value. The inset of (b) suggests that the fluctuations (scaled by system size) around the mean value will vanish in the thermodynamic limit. For visibility considerations, we used kernel density estimation in plotting the distributions.
  }
  \vspace{2.0mm}
  \label{fig:sawtooth_std}
\end{figure}

From the above discussion, we can identify two distinct dynamical regimes: i) In the absence (or weak rate) of measurements, the entanglement entropy is expected to grow linearly with time, bounded from above by half the system size. ii) Following the rare instance of strong measurements, we expect a sharp drop in the entanglement entropy. Combining the above two behaviors leads to the aforementioned sawtooth structure shown in \cref{subfig:sawtooth}. 
Given the above reasoning, one would expect a broad distribution of $S(t)$ resulting from the abrupt drop in values following strong measurements and the slow climb toward saturation during weak measurement periods, seen in \cref{fig:sawtooth_std} and contrasted with the standard volume law phase.
An interesting figure of merit for the broad distribution is the ratio between the standard deviation and the mean. While the standard volume law phase $\sigma(S)/\bar{S}$ tends to zero in the thermodynamic limit, for temporal randomness, they are of the same scale, see \cref{fig:sawtooth_std}.

\begin{figure}[h!]
\begin{center}
    \includegraphics[width=0.48\textwidth]{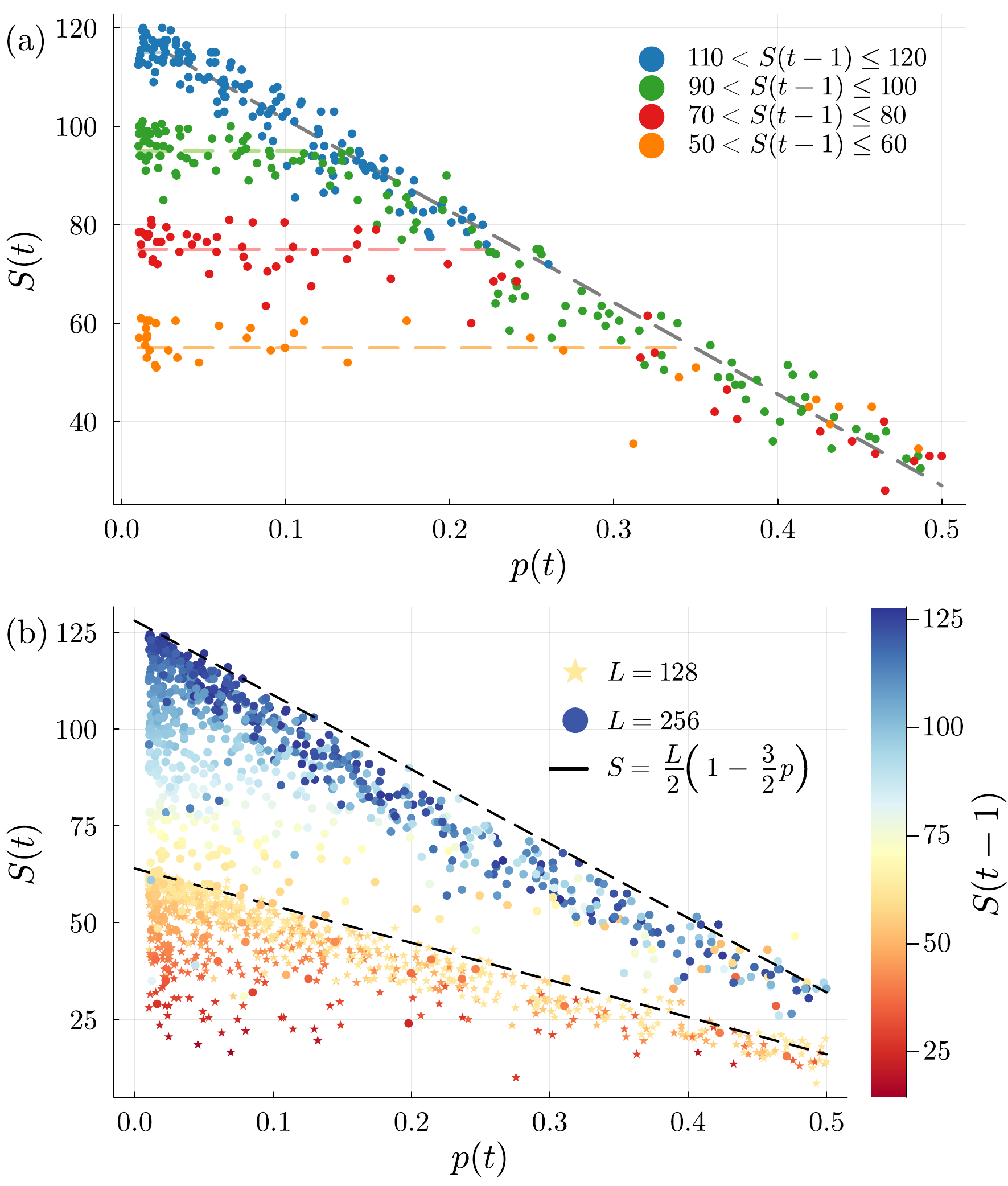}
    \end{center}
    \captionsetup[subfigure]{labelformat=empty}
    \subfloat[\label{subfig:Spost_Spre}]{}
    \subfloat[\label{subfig:Spost_tp}]{}
  \vspace{-3.0mm}
  \caption{(a) Scatter plot of $S(t)$ following an instance of measurement rate $p(t)$ for $L=256$. Different colors represent different ranges of $S(t-1)$. For sufficiently large $p(t)$, the spread of $S(t)$ weakly depends on $S(t-1)$. The dashed lines are guides to the eye and agree with the definition of $Q(L,p(t),S(t-1))$ described in the text. (b) The entanglement entropy immediately after a measurement $p(t)$. The color of the marker indicates the entanglement entropy value before this measurement. All data points were measured for $\bar{p}=0.005$.
  }
  \vspace{2.0mm}
  \label{fig:sawtooth_Spost}
\end{figure}

In the following, we construct a simple heuristic stochastic dynamical model that attempts to capture the sawtooth structure and the sublinear scaling. Explicitly, we suggest the following dynamics:
\begin{equation}
    S(t)=\begin{cases}
    \min(L/2,S(t-1)+1) & p(t)<p_0 \\
    Q(L,p(t),S(t-1)) & p(t)>p_0
    \end{cases}
    \label{eq:St_master}
\end{equation}
In the above equation, $p(t)$ is the measurement rate at time $t$, as defined above. The first line corresponds to the linear rise of entanglement entropy for a weak measurement rate, bounded from above by the saturation value. $p_0$ serves as an upper cutoff for the weak measurement regime. The second line corresponds to strong measurement events that produce a decrease of the entanglement entropy as captured by the random function $Q(L,p(t),S(t))$, which we estimate empirically below. 

In \cref{subfig:Spost_Spre}, we use a scatter plot to assess the values of $S(t)=Q(L,p(t),S(t-1))$, given $S(t-1)$ and $p(t)$ for $L=256$. We note that $Q(L,p(t),S(t-1))$ depends on $S(t)$ only as an upper bound. In particular, $S(t)$ for sufficiently large $p(t)$ is dominated by the values of $p(t)$ and displays a weak dependence on $S(t-1)$, as long as $S(t-1)>S(t)$. 

\begin{figure}[t!]
\begin{center}
    \includegraphics[width=0.48\textwidth]{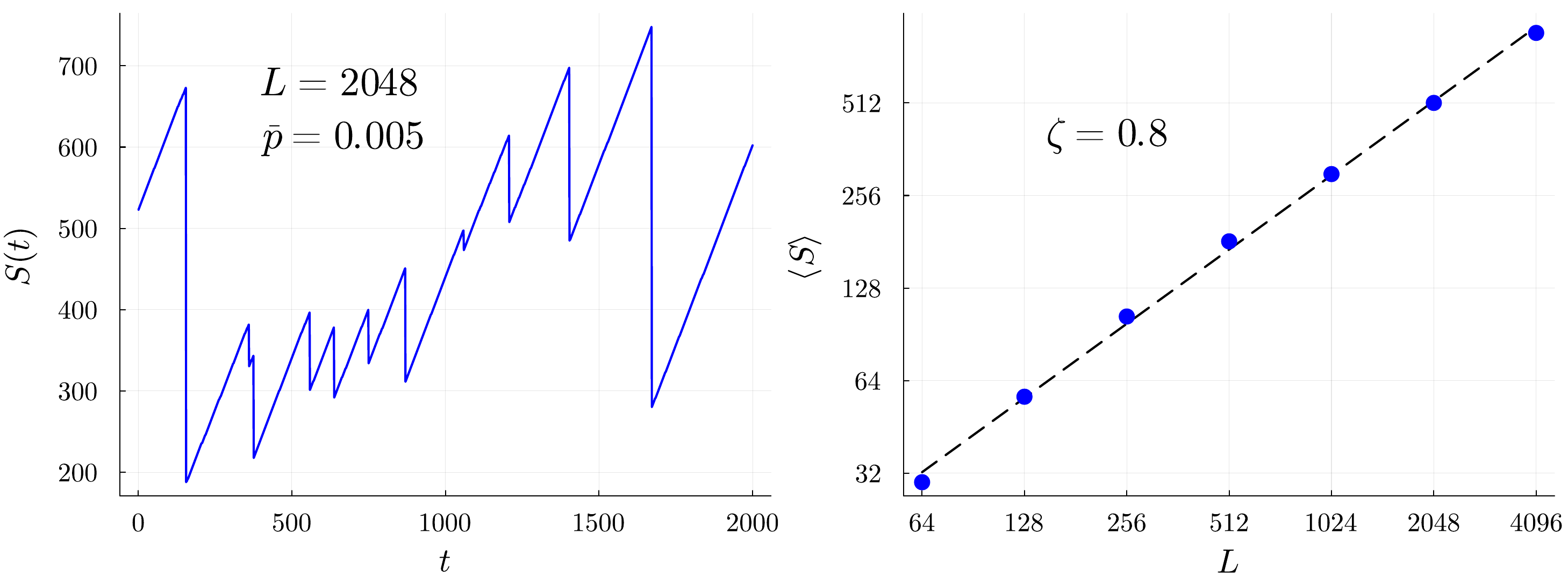}
  \end{center}
  \vspace{5.0mm}
  \caption{(Left) Sawtooth dynamics of $S(t)$ as generated by the toy model \cref{eq:St_master} (Right) $L$ scaling of the average $S_t$ at long times, for $\bar{p}=0.005$, and power-law in distribution $p[q]$ in \cref{eq:St_master}, which yields $\zeta=0.8$, in agreement with the Clifford circuit simulation.}
  \vspace{2.0mm}
  \label{fig:toymodel}
\end{figure}

To quantify this observation, we construct an effective data-based model, where we estimate the statistics of $S(t)$ values resulting from a single circuit layer with a specific $p(t)$ given $S(t-1)$ and the system size $L$. In \cref{subfig:Spost_tp}, we show that measuring at a measurement rate of $p$ results in an upper bound on the entanglement entropy immediately after this time instance to $g_{\text{\tiny upper}}(p,L)\coloneqq \max S\qty(t|p\qty(t)=p)= \frac{L}{2}\qty(1-\frac{3}{2}p)$. The average value of $S$ follows a similar relation $g_{\text{\tiny av}}(p,L)\coloneqq \left\langle S\qty(t|p\qty(t)=p)\right\rangle = \frac{4L}{9}\qty(1-\frac{3}{2}p)$. We utilize those results to construct a probability distribution function for $Q(L,p(t),S(t-1))=\min \qty(q,S(t-1))$. Our first choice for $P[q]\sim q^\gamma$ such that $q_{\max}=g_{\text{\tiny upper}}\qty(p\qty(t),L)$ and $\langle q\rangle =g_{\text{\tiny av}}\qty(p\qty(t),L)$, explicitly, $\gamma=\qty[g_{\text{\tiny upper}}\qty(p\qty(t),L)/g_{\text{\tiny av}}\qty(p\qty(t),L)]-1$. As a second choice, we take $P[q]$ to be normally distributed, with a mean $g_{\text{\tiny av}}(p,L)$ and standard deviation $g_{\text{\tiny std}}(p,L)=\frac{\sqrt{L}}{4}$ to match the data.  

We simulated \cref{eq:St_master}, with $\bar{p}=0.005$, and analyzed resulting stochastic dynamics. We found that with either choice of distributions, we successfully capture the sublinear growth of the entanglement entropy $S\sim L^\zeta$ with $\zeta\approx 0.8$ with power-law distribution and $\zeta\approx 0.75$ with normal distribution, see \cref{fig:toymodel}

We emphasize that our effective model qualitatively captures the dynamics of the full circuit dynamics only in the limit of small $\bar{p}$. 

\section{Dynamics of the entanglement entropy in the sub-volume phase}
\label{apdx:dynamical_exps}

\subsection{Temporal growth of the entanglement entropy}
\label{subapdx:kappa_supp}

\begin{figure}[tbh!]
\begin{center}
    \includegraphics[width=0.48\textwidth]{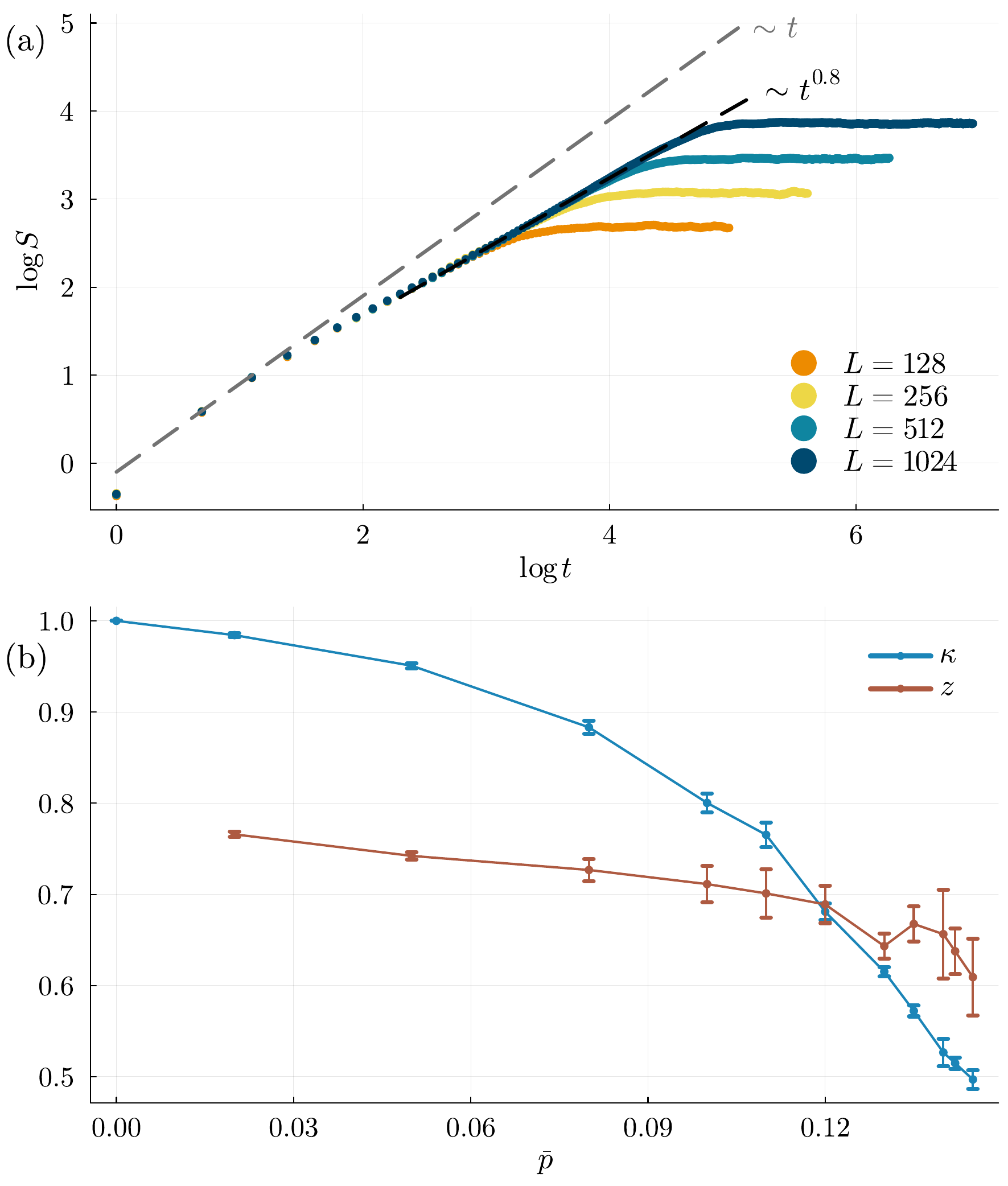}
  \end{center}
  \vspace{2.0mm}
  \captionsetup[subfigure]{labelformat=empty}
    \subfloat[\label{subfig:kappa}]{}
    \subfloat[\label{subfig:kappa_vs_p}]{}
  \caption{(a) The temporal evolution of the averaged entanglement entropy with $\bar{p}=0.1$. Different colors denote different system sizes. Within the studied range of system sizes, the pre-saturation growth is sublinear in time. For the presented $\bar{p}$, the exponent $\kappa\approx 0.8$, where $S\sim t^\kappa$. (b) The evolution of the exponent $\kappa$ (blue line), which gradually decreases from $\kappa=1$ for a unitary circuit to $\kappa\approx 1/2$ near $\bar{p}_c$. Similarly, the dynamical exponent $z$ (red line), defined via the saturation time $\tau_{\text{sat}}\sim L^z$, is less than unity throughout the entangling phase.}
  \label{fig:fig_kappa}
\end{figure}

\begin{figure*}[t!]
\vspace{0.0mm}
\begin{center}
    \includegraphics[width=1.0\textwidth]{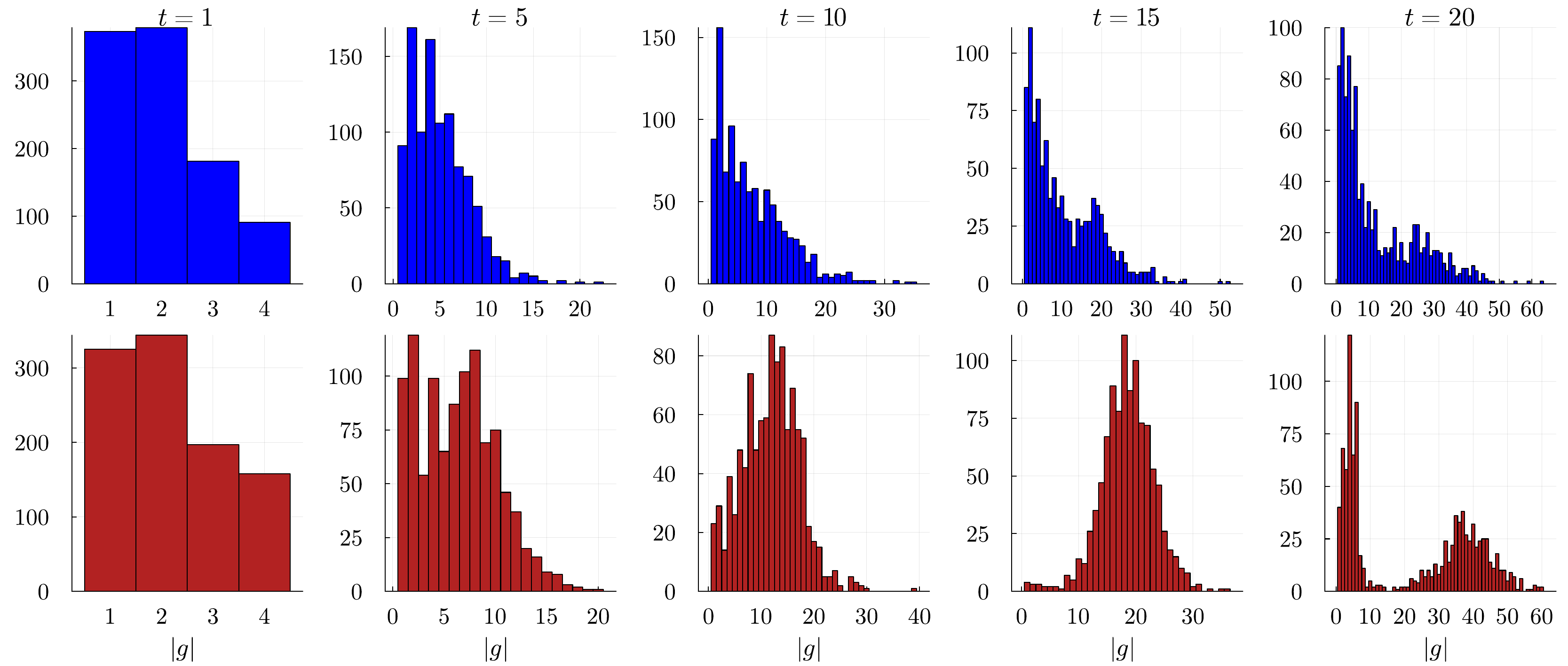}
  \end{center}
  \vspace{5.0mm}
  \caption{Evolution of stabilizer length distributions $|g|$ at selected time steps. Upper panels (blue) show the progression for a circuit with uniform measurement rate $p=0.1$, while lower panels (red) demonstrate the dynamics under temporally random measurements with an average rate $\bar{p}=0.1$.}
  \vspace{0.0mm}
  \label{fig:stab_dist_evol}
\end{figure*}

In this section, we study the real-time dynamics growth of the half-cut entanglement entropy. As with the sub-volume-law behavior, we allow for a power-law scaling deviating from linearity at times $t\ll L$,
\begin{equation}
    S(t,L)\sim t^{\kappa}.
\end{equation}
To extract the temporal exponent $\kappa$, we examine the growth of the averaged $S$ at early times, prior to saturation, see for example \cref{subfig:kappa}. Importantly, curves of different system sizes share the same trajectory before smaller systems reach saturation, and this shared window represents the thermodynamic value and is used to calculate $\kappa$. Within the entangling phase, $\kappa$ gradually decreases from $\kappa=1$ at the unitary point to $\kappa\approx 1/2$ near criticality, as shown in \cref{subfig:kappa_vs_p}. Further analysis of the pre-saturation dynamics via the lens of the clipped gauge stabilizer length statistics will follow immediately. 

This observation of nontrivial $\kappa$ in the entangling phase is surprising. In this regime, we do not anticipate that boundary conditions would affect the dynamics, and hence, predictions based on spacetime rotation are expected to hold. In particular, in the quenched disorder model, a volume-law phase $S\sim L$ was expected at low measurement rate, but not carefully tested \cite{Zabalo2022InfiniteRand}. 
$S\sim L$ would translate under spacetime rotation to linear-in-time growth of the entanglement entropy in the temporally random model. However, our numerical estimates suggest a sublinear-in-time growth. This discrepancy can result from a slow, finite-size flow of exponents beyond the reach of our numerical calculations. A more interesting possibility would be if this indicates true asymptotic deviation from volume-law entanglement scaling in the quenched disorder model and from linear-in-time entanglement growth in the temporally random model, thus telling us that these entangling phases are, in this sense, critical. We leave this open question to future studies.

\subsection{Pre-saturation dynamics of stabilizers length statistics}
\label{subapdx:clipped_gauge}

\begin{figure}[b!]
\vspace{-2.0mm}
\begin{center}
    \includegraphics[width=0.45\textwidth]{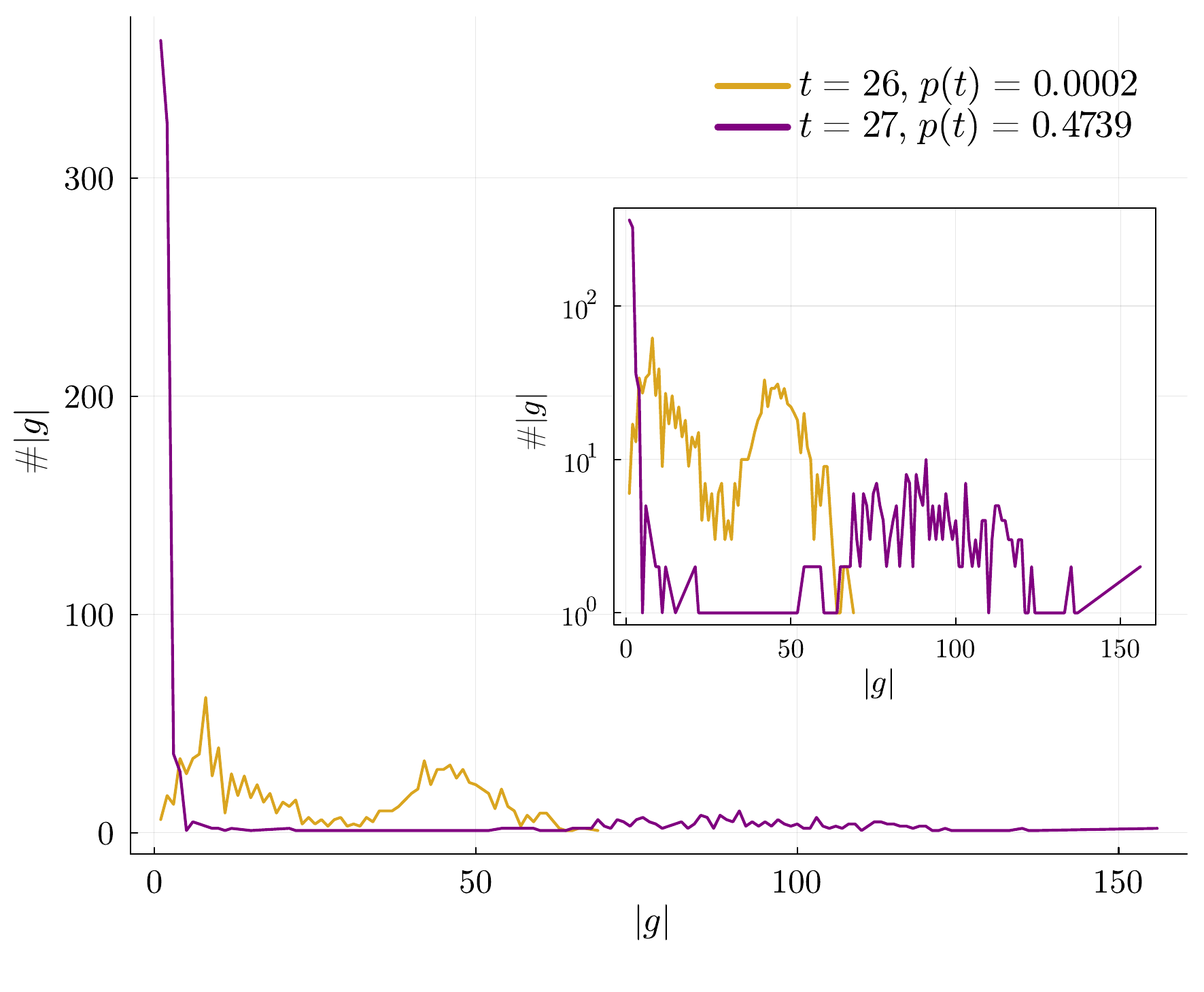}
  \end{center}
  \vspace{3.0mm}
  \caption{Impact of a strong measurement event ($p(t)=0.4739$) on stabilizer length distribution. The comparison of distributions immediately before and after the event illustrates that while most stabilizers are shortened, the remaining ones are enlarged due to teleportation.}
  \vspace{3.0mm}
  \label{fig:stab_dist_prepost}
\end{figure}

In this section, we examine the temporal evolution of the stabilizer length statistics within the clipped gauge framework \cite{Nahum2019MIPT,Fisher2019MIPT}. To that end, we simulate a sufficiently large system size comprising $L=1024$ qubits, allowing us to study the dynamics for broad pre-saturation time windows. Results are presented for open boundary conditions. We consider both uniform and temporally random measurement rates. Our focus is on the early stages of circuit evolution, for which hybrid circuit dynamics exhibit a linear-in-time entanglement entropy growth for the uniform case. Whereas the temporally random model demonstrates a potential sublinear growth $S(t)\sim t^{\kappa}$, with an exponent $\kappa\leq 1$ that varies with $\bar{p}$.

\begin{figure*}[t!]
\vspace{3.0mm}
\begin{center}
    \includegraphics[width=1.0\textwidth]{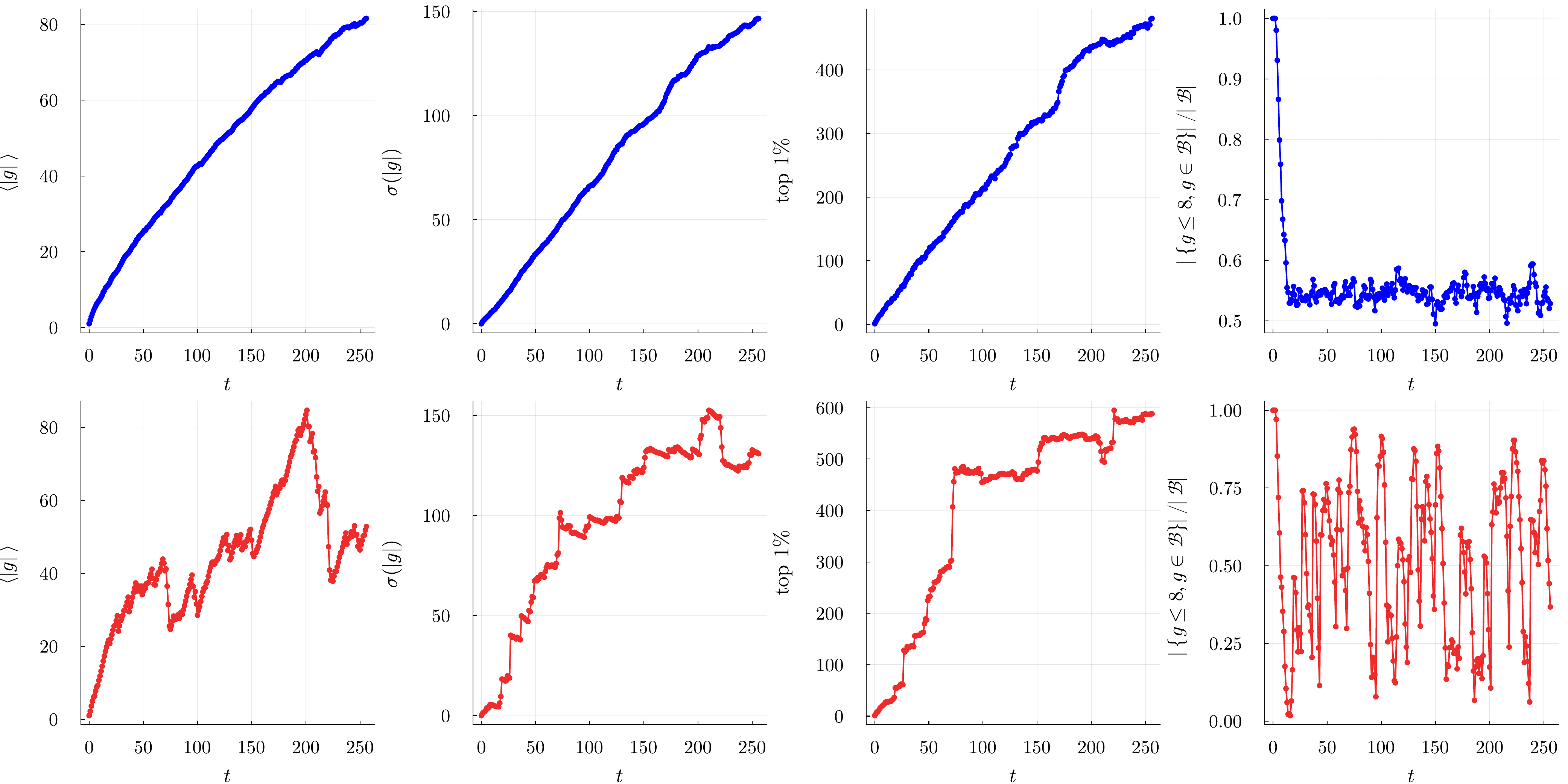}
  \end{center}
  \vspace{5.0mm}
  \caption{Temporal evolution of key statistical metrics for uniform measurement rate (upper row) and temporally random measurements (lower row), both with $\bar{p}=0.1$. From left to right -- first panel: mean stabilizer length; second panel: distribution standard deviation; third panel: the length of the stabilizer at the top 1\%; fourth panel: fraction of short stabilizers (length $\leq 8$ qubits). These metrics highlight the contrasting dynamics between uniform and random measurement protocols.}
  \label{fig:stab_stats}
\end{figure*}

In the clipped gauge, the entanglement entropy of a segment $A$ is defined by \cite{Ultrafast2023}
\begin{equation}
    S(A)=|A|-|\{g\in \mathcal{S} \,|\, \text{supp}(g)\subseteq A\}|,
\end{equation}
where $\mathcal{S}$ is a set of generators of the stabilizers group in the clipped gauge, and $\text{supp}(g)$ is the spatial support of the generator $g$ as defined by the clipped gauge conditions. With the above definition in mind, we can correlate the generator statistics and the entanglement entropy by noting that short stabilizers correspond to short-range entanglement (low entanglement entropy), whereas long stabilizers indicate long-range entanglement (high entanglement entropy).

\cref{fig:stab_dist_evol} tracks the temporal evolution of the distribution of stabilizer lengths in a single circuit realization. We compare a uniform measurement rate (upper panels, blue) and temporally random measurement rate (lower panels, red), both with $\bar{p}=0.1< \bar{p}_c$. We observe that the standard uniform model develops a concentration of short stabilizers associated with measurement events and a gradually increasing tail of long stabilizers generated primarily by unitary dynamics. By contrast, the temporally random model alternates between qualitatively distinct forms: 1) Time periods characterized by weak $p(t)$ (nearly unitary evolution) for which the distribution is concentrated at a finite typical stabilizer size with a growing in time mean value. 2) Post strong measurement events, which dramatically shorten most stabilizers while simultaneously enlarging a fraction of stabilizers via teleportation. A concrete example of the latter is demonstrated in \cref{fig:stab_dist_prepost}.

The above evolution patterns are also revealed by examining several statistical metrics of the stabilizer length distribution, such as the mean, standard deviation, and the proportion of short stabilizers, as analyzed in \cref{fig:stab_stats}. The uniform $p$ model exhibits smooth, monotonic growth in both the mean length and the standard deviation ($\sigma(|g|)\sim t$), with rapid convergence in the fraction of short stabilizers. The temporally random model, however, demonstrates notably different behavior, characterized by irregular, non-monotonic evolution in all these metrics driven by rare strong measurement events.

\subsection{Evolution of dynamical exponent $z$ in the sub-volume-law phase}
\label{subapdx:z_exp}

The unconventional spatial and temporal power-law scaling exponents suggest a nontrivial dynamical exponent, $z$, relating space and time length scales. To compute $z$, we primarily focus on the entanglement entropy saturation time. This is achieved by initializing a random product state and tracking the evolution of the entanglement entropy for varying system sizes. As explained before, in the thermodynamic limit $L\to\infty$, the entanglement entropy follows a power-law growth as $S\sim t^\kappa$. This is exemplified in \cref{subfig:Sdyn_extract}, where the pre-saturation evolution of curves corresponding to an increasing range of system sizes converges to the thermodynamic limit. The saturation time is then defined as the time at which the entanglement entropy in the thermodynamic limit reaches its saturation value of a given finite-size system, $S_{\text{sat}}(L,t\to \infty)$. The dynamical exponent $z$, is then evaluated by a a fit to the standard space-time scaling form. $\tau_{\text{sat}}\sim L^z$ \cite{Zabalo2022InfiniteRand,Shkolnik2023QP}, see \cref{subfig:tsat_vs_L}.

\begin{figure}[h!]
\begin{center}
    \includegraphics[width=0.48\textwidth]{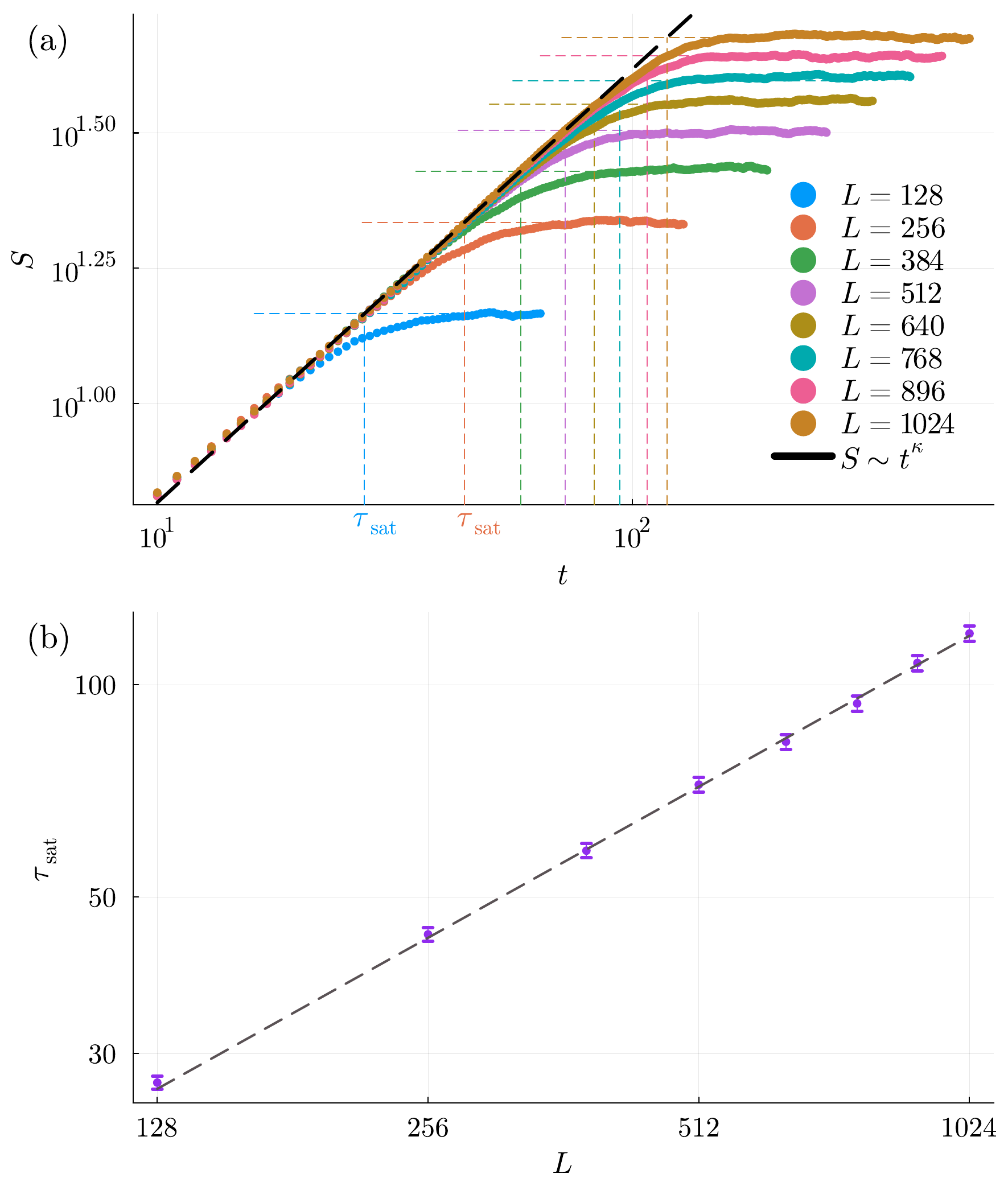}
  \end{center}
  \vspace{4.0mm}
  \captionsetup[subfigure]{labelformat=empty}
    \subfloat[\label{subfig:Sdyn_extract}]{}
    \subfloat[\label{subfig:tsat_vs_L}]{}
  \caption{(a) The temporal evolution of the entanglement entropy for $\bar{p}=0.1$. To evaluate the saturation time, we first ensure convergence to the thermodynamics limit, which follows a power-law scaling form $S\sim t^\kappa$, with $\kappa=0.80(2)$. Then, we project the saturation value of the entanglement entropy at a given system size $L$ to the infinite system size curve. The intersection point sets the saturation time $\tau_{\text{sat}}(L)$. (b) An example for the computation of the exponent $z$ via the saturation time $\tau_{\text{sat}}$ scaling as a function of system size for $\bar{p}=0.1$.}
  \vspace{2.0mm}
  \label{fig:Sdyn_extract}
\end{figure}

To support the above calculation, we also compute typical time scales via two additional approaches. The first observable is the relaxation time extracted from the temporal autocorrelation function of the entanglement entropy, evaluated at late times, beyond the averaged saturation time. Explicitly, 
\begin{equation}
    R_S(\tau)=\langle S(t)S(t+\tau)\rangle_{t\gg \tau_{\text{sat}}}-\langle S(t)\rangle ^2.
\end{equation}
Assuming an exponential decay, we extract the corresponding decay factor $T$ for a given system size $L$ \cite{to_be_published}. The second is the entanglement time $t_{\text{ent}}$ associated with the tripartite mutual information defined in the main text. The above analysis is summarized in \cref{subfig:z_3methods}. Importantly, all methods yield approximately the same $z(\bar{p})$. 

Interestingly, the dynamical exponent $z$ is directly related to the exponents pair $\kappa$ and $\zeta$. From the above discussion, it follows that $S_{\text{sat}}(L)\sim \tau_{\text{sat}}^\kappa(L)\sim L^\zeta$, such that $\tau_{\text{sat}}\sim L^{\zeta/\kappa}$. Consequently, we identify 
\begin{equation}
    z= \zeta/\kappa.
\label{eq:z_zeta_kappa}
\end{equation}
We numerically verify this relation explicitly in \cref{subfig:zeta_comp}. A key observation is that in the sub-volume-law phase, $\zeta(\bar{p})<\kappa(\bar{p})\leq 1$ for all $\bar{p}$, which, when combined with \cref{eq:z_zeta_kappa}, implies that $z$ is less than unity across the entire phase, as depicted in \cref{subfig:kappa_vs_p}. This allows us to conclude that dynamics, throughout the entangling phase, is dominated by temporal Griffiths effects.

\begin{figure}[h!]
\begin{center}
    \includegraphics[width=0.48\textwidth]{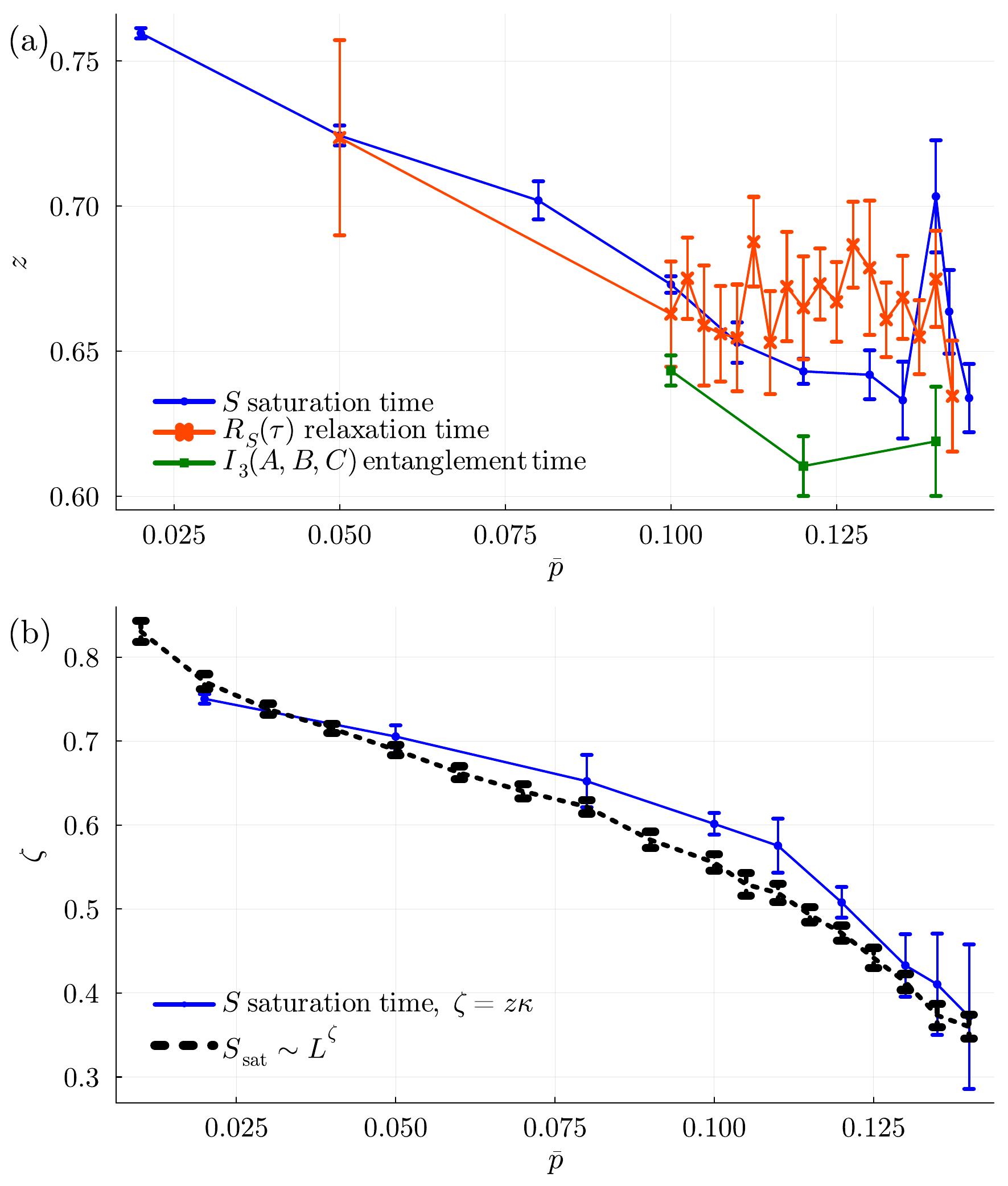}
  \end{center}
  \vspace{0.0mm}
  \captionsetup[subfigure]{labelformat=empty}
    \subfloat[\label{subfig:z_3methods}]{}
    \subfloat[\label{subfig:zeta_comp}]{}
  \caption{(a) the dynamical exponent $z$ as was calculated using the entanglement entropy growth (in blue), the autocorrelation relaxation time (in red) and the entanglement time (in green). (b) The values of $\zeta$ from \cref{subfig:zeta_grif} against the product of $z$ and $\kappa$ calculated from the space and time scaling of $S$ (in blue).}
  \vspace{0.0mm}
  \label{fig:z_kappa}
\end{figure}

\section{Statistics of $\mathcal{I}_3$}
\label{apdx:I3_dist}
In the main text, we employed the observable $\ptmi$ to detect the critical point and extract critical exponents. This choice allows for the taming of strong finite-size effects associated with temporal rare regions that lead to a broad $\mathcal{I}_3$ distribution.  In \cref{fig:I3_dist}, we depict histograms of $\mathcal{I}_3$ in various regimes, which indeed display a broad distribution, motivating the focus on the finite-size analysis of $\ptmi$. The different tendencies of $\ptmi$ between the two phases and the plateau near criticality are manifest in the scaling across a wide region of $\bar{p}$ values, which we used in \cref{subfig:ptmi_collapse}.

\begin{figure}[h!]
\begin{center}
    \includegraphics[width=0.48\textwidth]{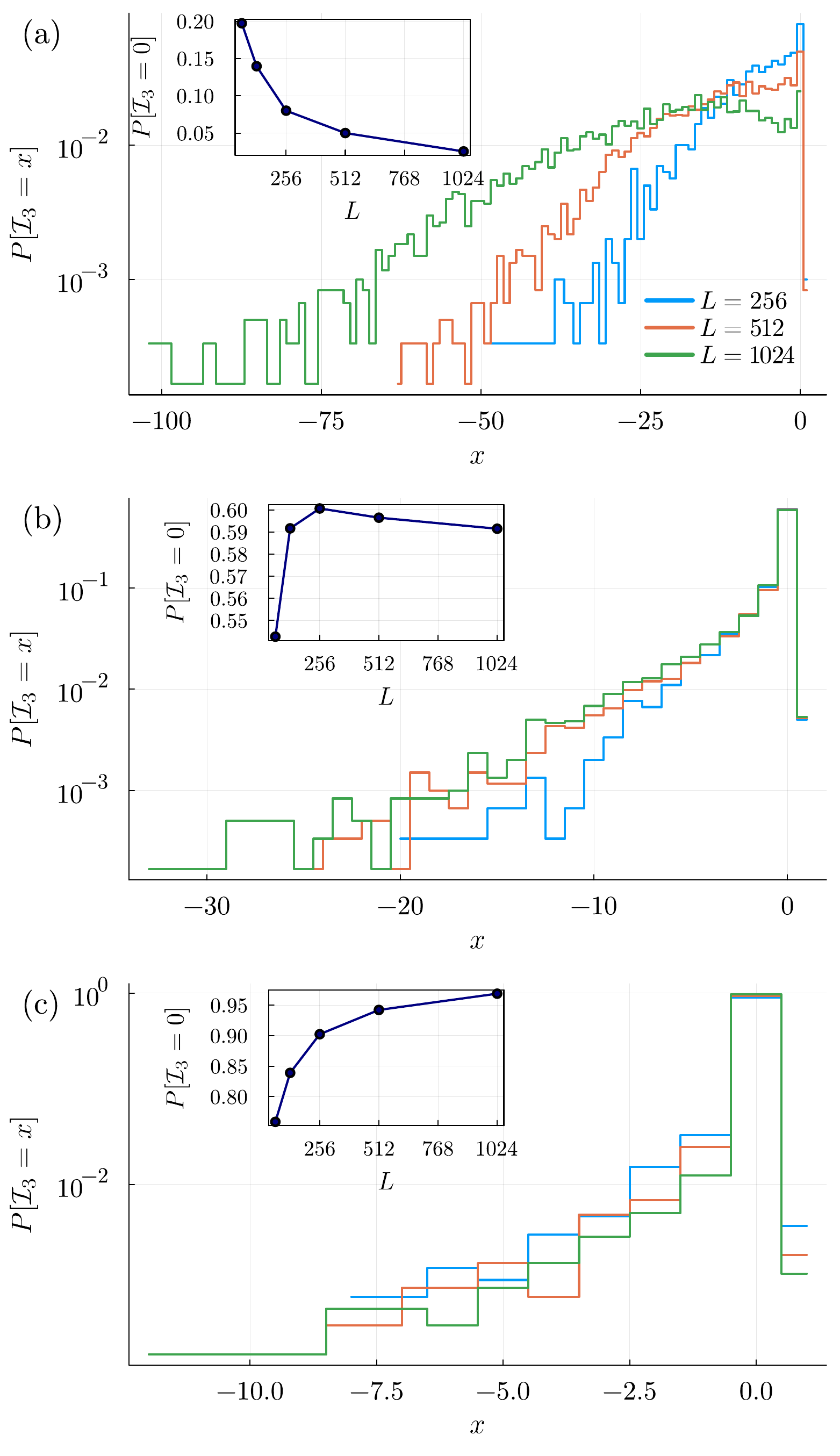}
  \end{center}
  \vspace{5.0mm}
  \caption{Histograms of $\mathcal{I}_3$ at $\bar{p}=0.1$, in the sub-volume-law phase (a), $\bar{p}\approx \bar{p}_c=0.15$ (b) and at $\bar{p}=0.18$ in the area-law phase (c). The insets finite scaling of  $\ptmi$ in the different phases.}
  \label{fig:I3_dist}
  \vspace{0.0mm}
\end{figure}

\section{Detailed analysis of critical data extraction via finite-size scaling}
\label{apdx:criticality}

This section offers a detailed analysis of the numerical methods employed to determine the critical point and related universal critical exponents. An accurate method for extracting the critical point and critical exponents is considering pairs of system sizes $L_<$ and $2L_<$. For each pair, we find the intersection of $\ptmi$ curves, which set the finite-size value of the critical point $\bar{p}_c\qty(L_<)$. In addition, the derivate 
\begin{equation}
    g(L_<)=\frac{d\ptmi(\bar{p},L)}{d \bar{p}}|_{\bar{p}_c\qty(L_<)}
\end{equation}
with respect to a deviation from criticality is anticipated to follow the scaling $g(L)=\qty(\log L)^{1/(\nu_{\tau} \psi_{\tau})}f'(0)$, according to our ansatz in \cref{eqn:ptmi_ansatz}.
In \cref{fig:pc_der_method}, we summarize the above analysis. We indeed find a convergence to the critical measurement rate $p_c=0.150(3)$ as a function of system size and an excellent agreement with the scaling ansatz, allowing for an accurate estimation of exponents product $\nu_\tau \psi_\tau$.

\begin{figure}[tbh!]
\begin{center}
    \includegraphics[width=0.48\textwidth]{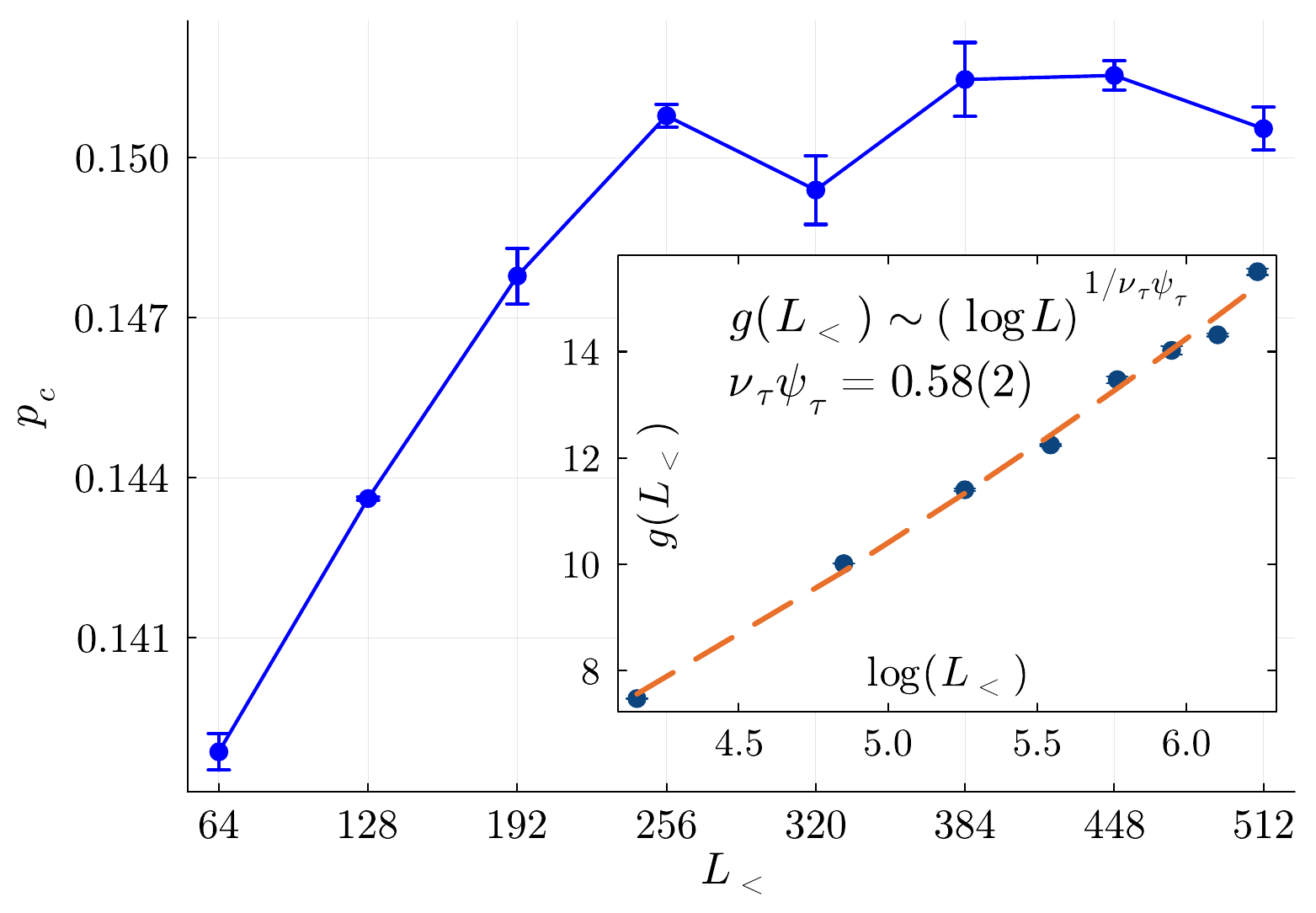}
  \end{center}
  \vspace{5.0mm}
  \caption{The critical point as evaluated using the intersection of pairs of $\ptmi$ curves. The $\bar{p}_c$ value converges to $\bar{p}=0.150(3)$. The inset depicts the growth of the derivative of $\ptmi$ at the critical point as $\qty(\log L)^{1/(\nu_{\tau}\psi_{\tau})}$ with $\nu_{\tau}\psi_{\tau}=0.58(2)$.}
  \vspace{2.0mm}
  \label{fig:pc_der_method}
\end{figure}

Next, we compute the entanglement entropy scaling function at long times, extending the critical scaling form $S(\bar{p}_c)\sim \qty(\log L)^{1/2\psi_{\tau}}$, as verified in \cref{subfig:Spc_vs_logL}. For small deviations from criticality, we expect the generalized scaling form to hold,
\begin{equation}
    S(p,L)=\qty(\log L)^{1/2\psi_{\tau}}f\qty[\qty(\bar{p}-\bar{p}_c)\qty(\log L)^{1/(\nu_{\tau}\psi_{\tau})}].
\label{eqn:S_ansatz}
\end{equation}
Utilizing the previously calculated values $\psi_{\tau}\approx0.3$ and $\nu_{\tau}\psi_{\tau}\approx 0.6$, without further tuning parameters, indeed yields a Widom scaling curve collapse, presented in \cref{fig:S_collapse}.

\begin{figure}[b!]
\begin{center}
    \includegraphics[width=0.48\textwidth]{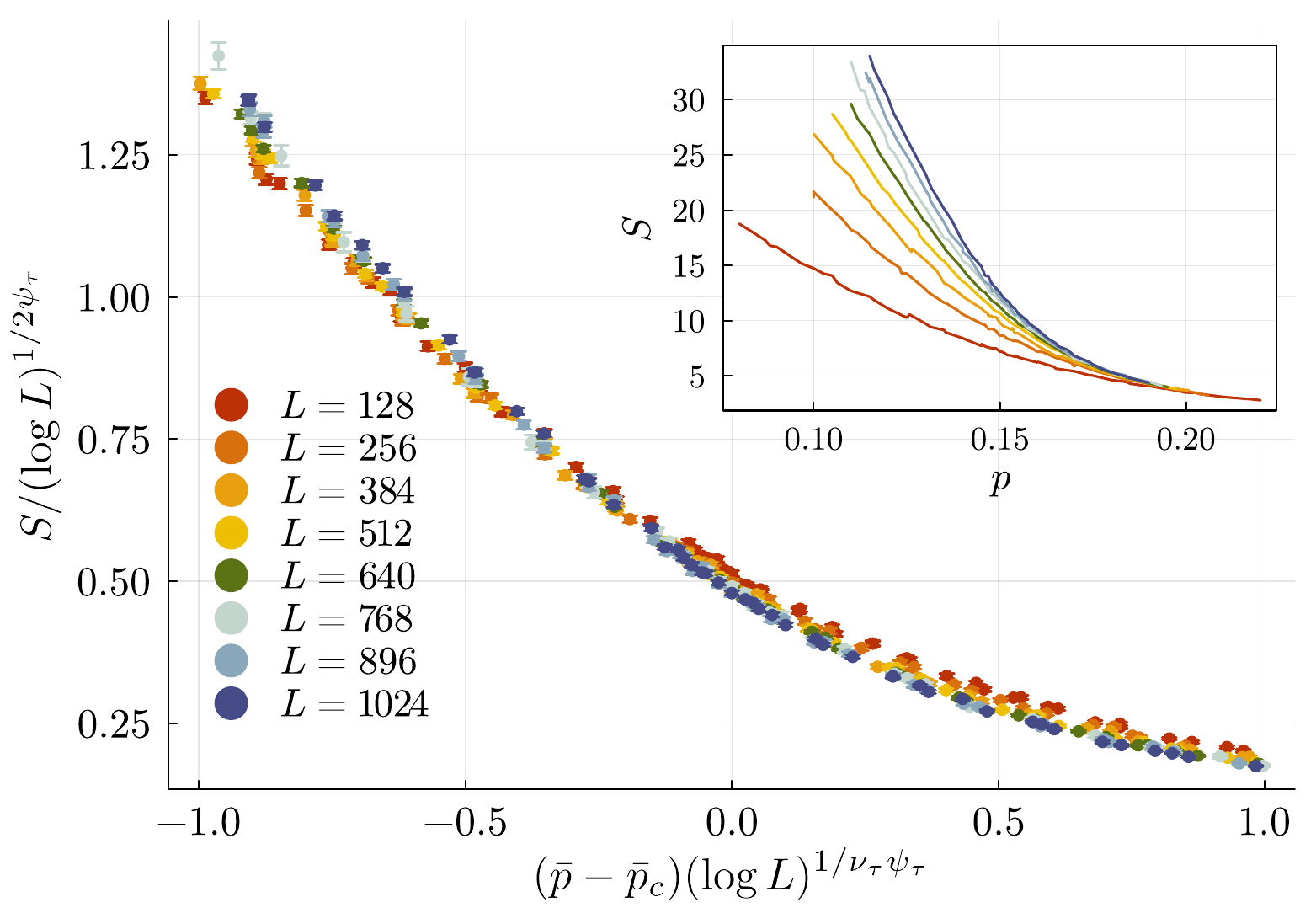}
  \end{center}
  \vspace{5.0mm}
  \caption{Widom scaling of the bipartite entanglement entropy with the pre-calculated values of $\psi_{\tau}$ and $\nu_{\tau}\psi_{\tau}$. The inset shows the data before scaling.}
  \vspace{0.0mm}
  \label{fig:S_collapse}
\end{figure}

Lastly, to rule out a power-law space-time scaling relations at criticality, we attempt to fit $P\qty[\mathcal{I}_3=0]$ to the standard power-law ansatz $P\qty[\mathcal{I}_3=0]\sim f\qty[\qty(\bar{p}-\bar{p}_c)L^{1/\nu}]$ for a sequence of increasing system sizes ranges, starting with $L_<$. As shown in \cref{fig:dec_powers}, we find a running in system size exponent, suggesting an RG flow to a vanishing exponent consistent with activated logarithmic scaling as our ultrafast ansatz.

\begin{figure}[ht!]
\vspace{-2.0mm}
\begin{center}
    \includegraphics[width=0.45\textwidth]{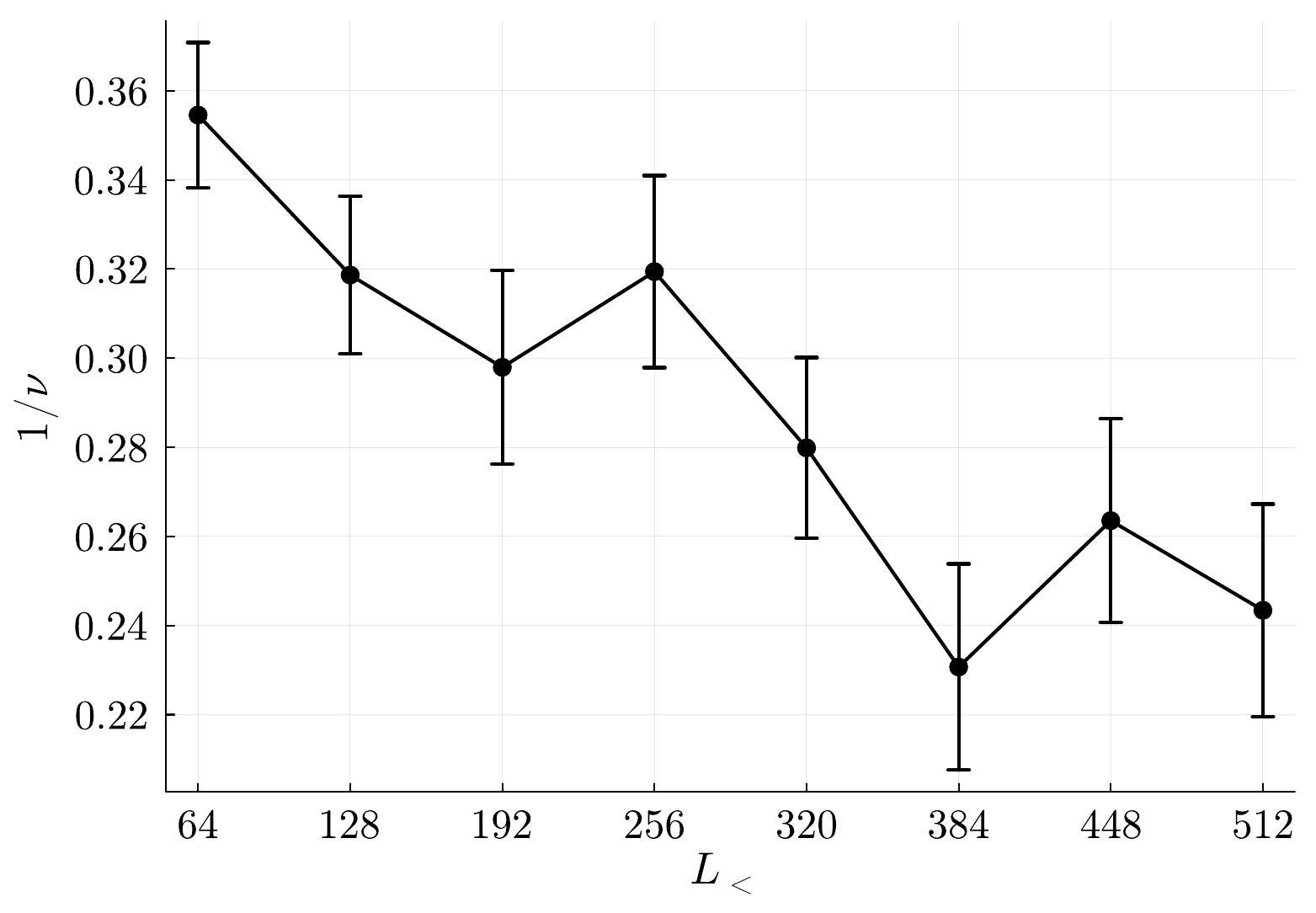}
  \end{center}
  \vspace{5.0mm}
  \caption{The correlation length critical exponent as extracted from the standard ansatz $P\qty[\mathcal{I}_3=0]\sim f\qty[\qty(\bar{p}-\bar{p}_c)L^{1/\nu}]$ decreases with the increasing system size.}
  \label{fig:dec_powers}
\end{figure}

\section{Spacetime-rotated Clifford circuit dynamics with quenched disorder}
\label{apdx:direct_rot}

\begin{figure}[t!]
\vspace{0.0mm}
\begin{center}
    \includegraphics[width=0.38\textwidth, angle=270]{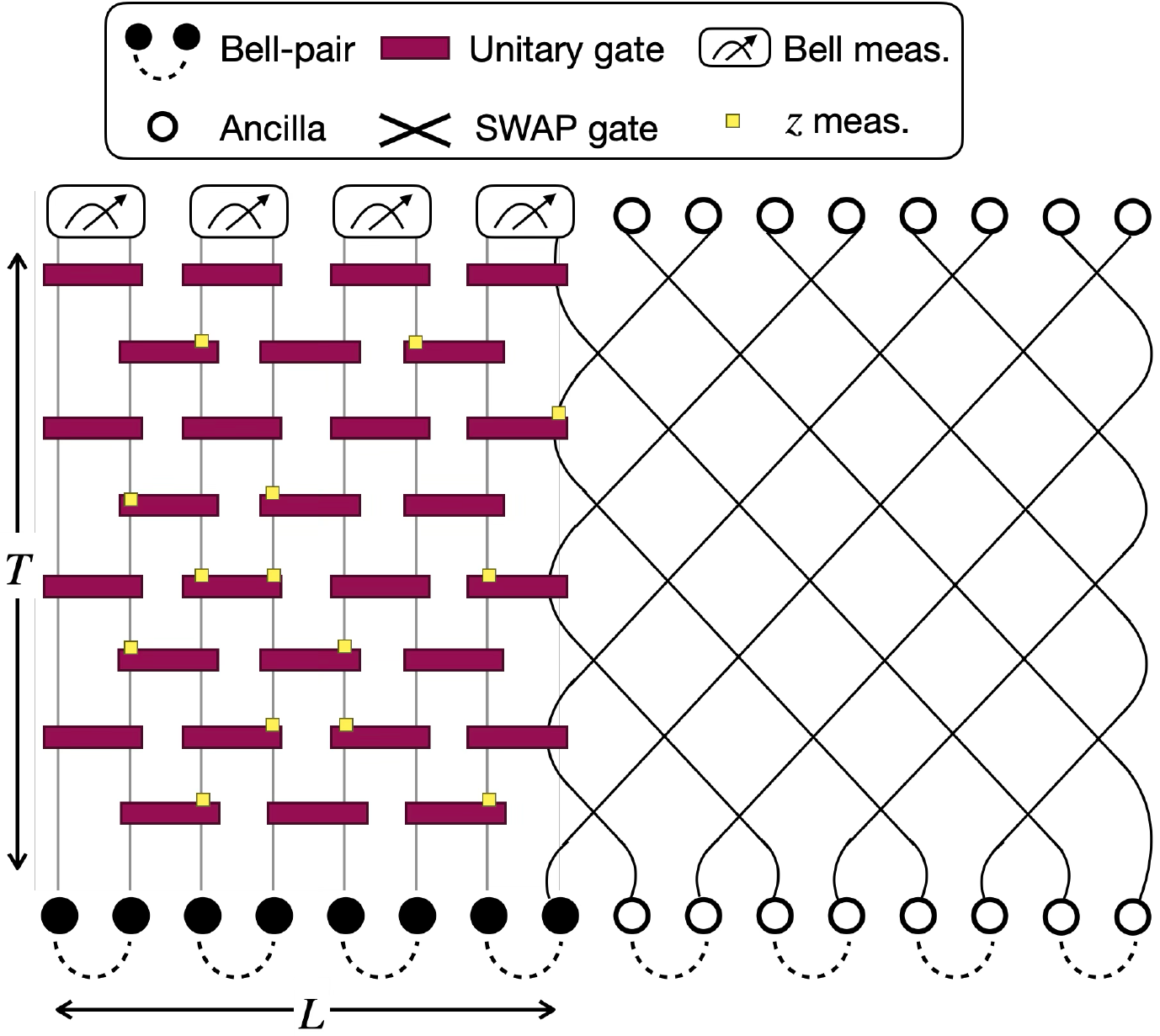}
  \end{center}
    \captionsetup[subfigure]{labelformat=empty}
    \subfloat[]{}
  \caption{We simulate a brickwork Clifford circuit of $L$ qubits for a time $T$ (each time step counts for both the even and odd two-qubit unitary gates layers). The qubits are measured with probability $p(x)$ that is space-dependent. The circuit is concatenated with $2T$ ancilla qubits subjected to only SWAP gates over nearest-neighboring ancillas, including the rightmost qubit in the original circuit. The SWAP gates project the state along the temporal axis to a spatial state over the ancillas at final time. The qubits are initially set in Bell-pairs and the $L$ original qubits are measured in the Bell basis with forced outcome at the end of the circuit.}
  \label{fig:sideways_sketch}
\end{figure}

In this section, we study a spacetime-rotated version of the infinite-randomness circuit model studied in Ref.~\cite{Zabalo2022InfiniteRand}. Since we do not consider dual-unitary dynamics, this model at the microscopic level is distinct from the one studied in the main text. Nevertheless, in the long-wavelength limit, this approach can potentially provide an alternative route for realizing ultrafast critical dynamics. If true, it allows generalizing the universality class of the temporally random model presented in the main text to a spacetime rotation of the infinite-randomness fixed point.

Our numerical implementation closely follows the spacetime-rotated circuit contraction introduced in \cite{PhysRevLett.126.060501,PhysRevX.12.011045}. We introduce \textit{quenched} randomness in measurement rates similarly to the main text and Ref.~\cite{Zabalo2022InfiniteRand} using $p(x) = \frac{1}{2}r^n$. The setup of the contracted circuit is depicted in \cref{fig:sideways_sketch}.

As evident from the evolution of the entanglement entropy, depicted in \cref{fig:direct_rot_EE}, we observe a MIPT with the increase in $\bar{p}$, in the spacetime-rotated model. The entangling phase, as expected from the above results and Ref.~\cite{PhysRevX.12.011045}, is characterized by a sub-volume-law growth of the entangled entropy (\cref{eqn:subvolumelaw}) with the exponent $\zeta$ decreasing upon approaching the phase transition, see the inset of \cref{fig:direct_rot_EE}.

\begin{figure}[b!]
\vspace{0.0mm}
\begin{center}
    \includegraphics[width=0.45\textwidth]{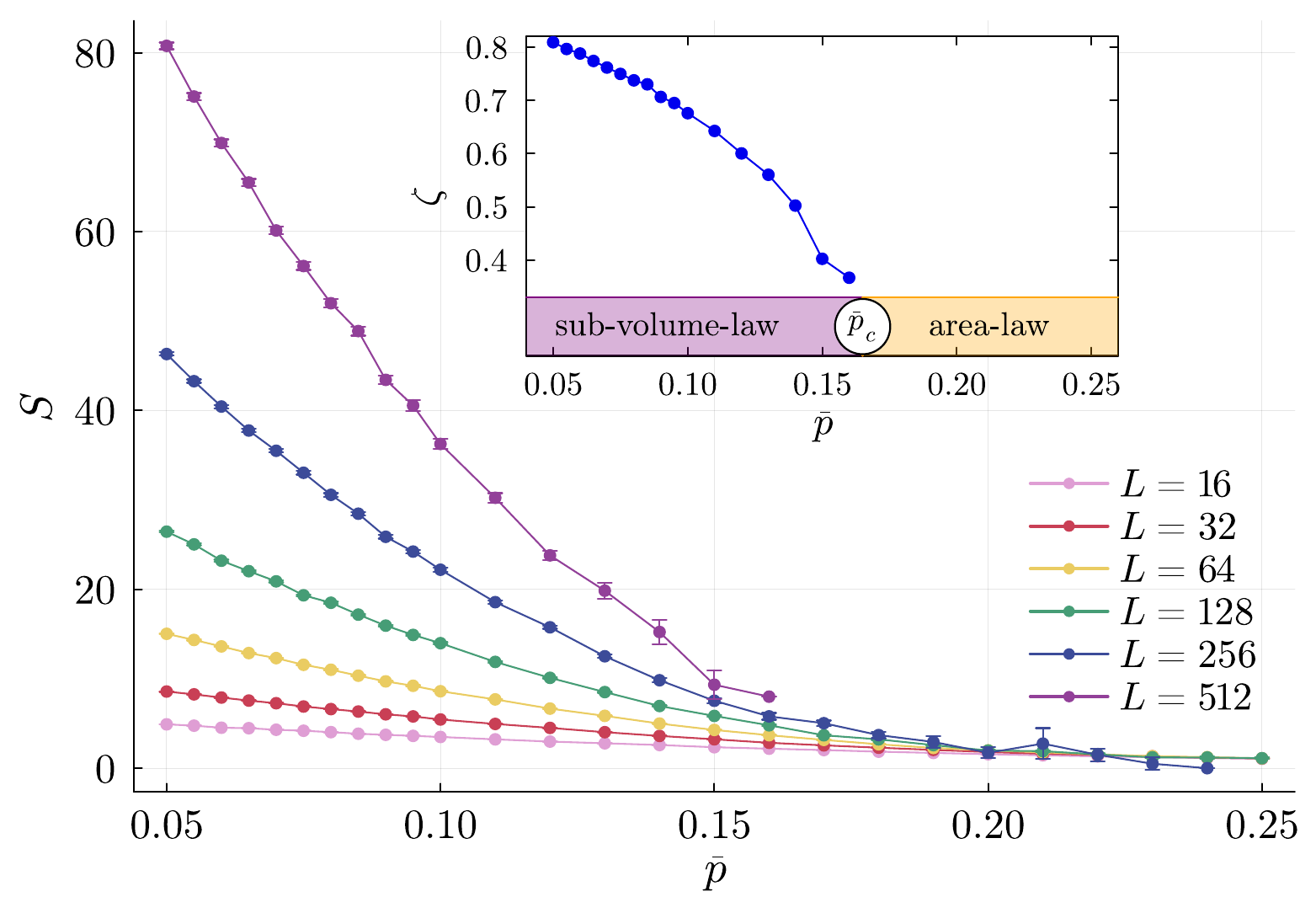}
  \end{center}
    \captionsetup[subfigure]{labelformat=empty}
    \subfloat[]{}
  \caption{The half-cut bipartite entanglement entropy of different system sizes as a function of the average measurement rate. We observe a phase transition between a sub-volume-law phase with a decreasing $\zeta$ exponent and a disentangling phase, as depicted in the inset.}
  \label{fig:direct_rot_EE}
\end{figure}

We follow to test the validity of our space-time scaling ansatz \cref{eqn:ptmi_ansatz} and \cref{eqn:S_scaling}, using the critical exponent values estimated using the temporally random model in the main text. A curve collapse analysis of the half-cut bipartite entanglement entropy (in systems of up to $L=128$), presented in \cref{fig:direct_rot_scaled_EE}, shows that this direct rotation is consistent with the results of the model studied in the main text. Specifically, the calculated critical exponents are $\psi_{\tau}=0.26(5)$ and $\nu_{\tau}=2$ with $\bar{p}_c=0.16(1)$. These values are close to the critical exponents that we concluded in the main text, suggesting that the difference between the naive expectation of $\psi_{\tau}=1/2$ and the calculated $\psi_{\tau}\approx0.3$ is inherent and not a mere finite-size effect. However, we emphasize that the resolution of our data from the rotated circuit may be insufficient (mainly due to small system sizes resulting from the postselection process) to accurately compute $\psi_{\tau}$ in the thermodynamic limit. Still, importantly, the ultrafast ansatz holds nonetheless.

\begin{figure}[ht!]
\vspace{0.0mm}
\begin{center}
    \includegraphics[width=0.45\textwidth]{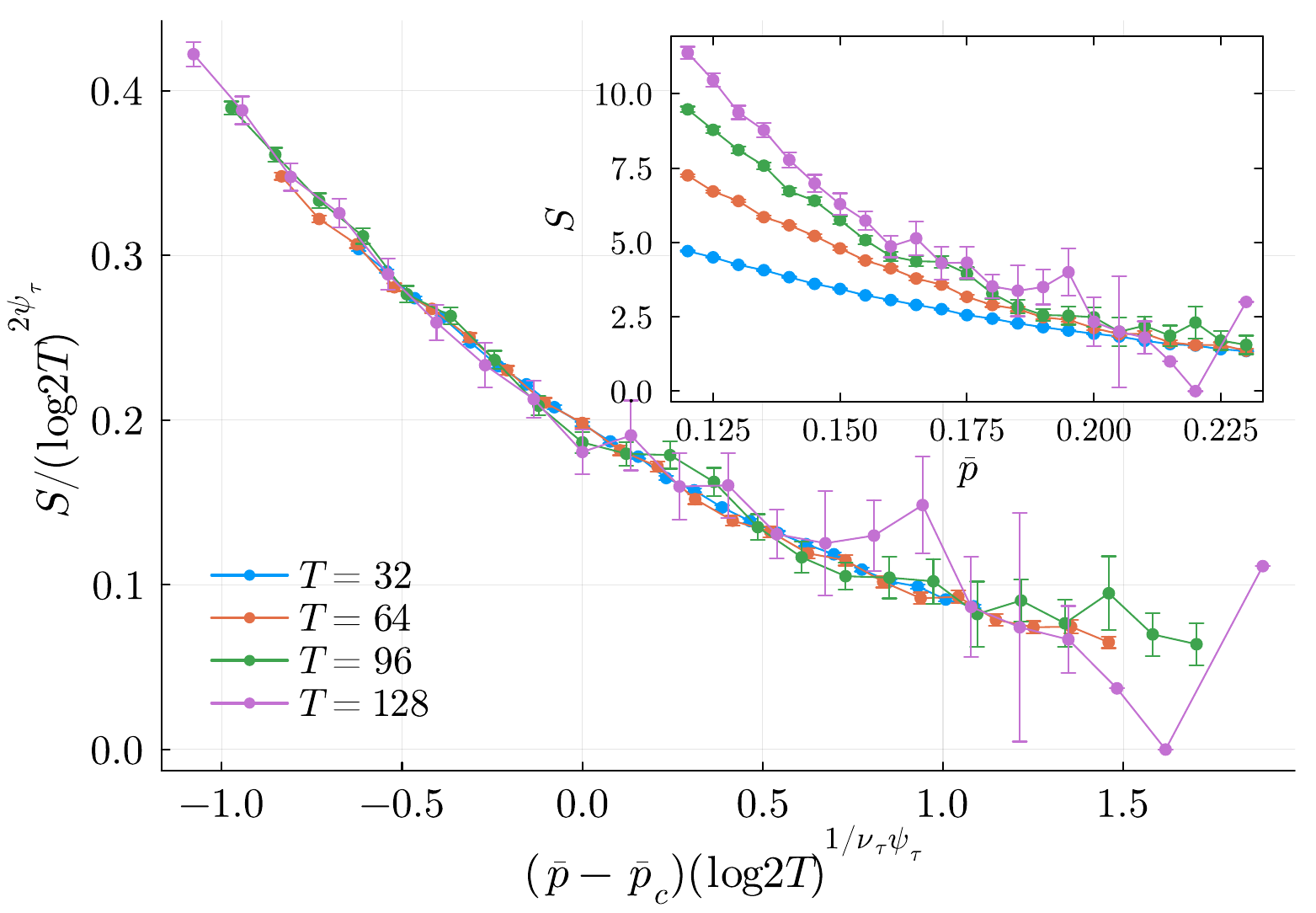}
  \end{center}
    \captionsetup[subfigure]{labelformat=empty}
    \subfloat[]{}
  \caption{Widom scaling of the half-cut bipartite entanglement entropy according to the ansatz $S\sim \qty(\log 2T)^{1/2\psi_{\tau}}f\qty[(\bar{p}-\bar{p}_c)\qty(\log 2T)^{1/\nu_{\tau}\psi_{\tau}}]$ with $\nu_{\tau}=2$, $\psi_{\tau}=0.26$ and $\bar{p}_c=0.16$. The inset shows the unscaled data.}
  \label{fig:direct_rot_scaled_EE}
\end{figure}

\section{Teleportation events in the information propagation}
\label{apdx:xI_teleportation}

As discussed in the main text, measurements can induce teleportation, which expands the range of entanglement. In the context of the information propagation metric $x_I$, such teleportation events would immediately increase the distance the information travels. The essence of teleportation is manifested in the averaged $x_I$ quantity by the superluminal dynamics, which could not have occurred by merely unitary dynamics that only allows an order-one increase in $x_I$ at each time step. Here, we demonstrate teleportation by examining the temporal trajectory of $x_I$ in a single circuit realization. This effect is directly shown in \cref{subfig:xI_jump}, where three different realizations of a system of $L=1025$ qubits in the entangling phase ($\bar{p}=0.05$) exhibit a remarkable jump in $x_I$ of more than 200 qubits -- nearly a fifth of the system size -- after just four time steps. Such a dramatic leap in information propagation is attributed to the action of strong measurements. In \cref{subfig:dxI_uni_rand} we compare the distribution of maximal $x_I$ jumps between uniform and temporally random measurement models at their corresponding critical points $p_c^{\text{\tiny uni}}\cong 0.16$ and $p_c^{\text{\tiny rand}}\cong 0.15$. While both models demonstrate teleportation events, the temporally random circuit typically achieves greater maximum distances, highlighting specific events for which $p(t)>\bar{p}_c$.

\begin{figure}[h!]
\vspace{0.0mm}
\begin{center}
    \includegraphics[width=0.45\textwidth]{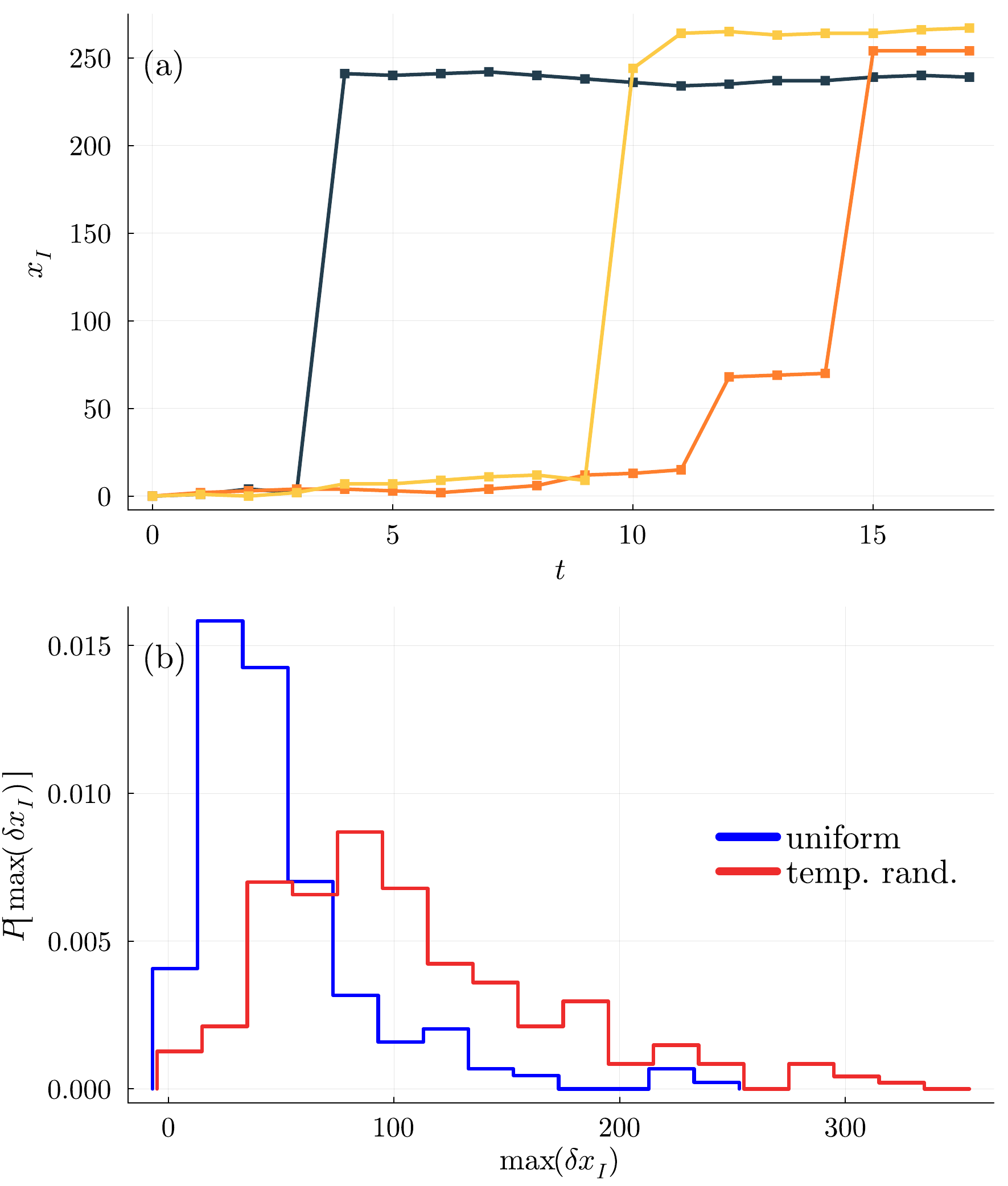}
  \end{center}
    \captionsetup[subfigure]{labelformat=empty}
    \subfloat[\label{subfig:xI_jump}]{}
    \subfloat[\label{subfig:dxI_uni_rand}]{}
  \caption{(a) Measurement-induced teleportation demonstrated through the information propagation $x_I$ in specific circuit realizations with $L=1025,\,\bar{p}=0.05$. Each color represents a different sample. (b) The probability distribution of the maximum single-step increase in $x_I$ for uniform (blue) and temporally random (red) measurement patterns at their critical point $p_c^{\text{\tiny uni}}\cong 0.16$ and $p_c^{\text{\tiny rand}}\cong 0.15$. Importantly, we only consider samples in which the ancilla survived longer than 32 time steps.}
  \label{fig:xI_teleportation}
\end{figure}

\pagebreak
\bibliography{ultrafast_mipt}

\end{document}